\pdfoutput=1
\documentclass[a4paper,american,floatfix,pdftex,superscriptaddress,%
aps,pra,%
reprint,twocolumn,final
]{revtex4-2}
\usepackage{amssymb}
\usepackage{amsmath}
\usepackage[T1]{fontenc}
\usepackage{graphicx}
\usepackage[utf8]{inputenc}
\usepackage[ams,todos]{CMImacros}
\usepackage{xr-hyper} 
\graphicspath{{./fig/}} 
\newcommand{\cfeldesy}{\affiliation{Center for Free-Electron Laser Science CFEL, Deutsches
      Elektronen-Synchrotron DESY, Notkestr. 85, 22607 Hamburg, Germany}}%
\newcommand{\uhhcui}{\affiliation{Center for Ultrafast Imaging, Universität Hamburg, Luruper
      Chaussee 149, 22761 Hamburg, Germany}}%
\newcommand{\uhhphys}{\affiliation{Department of Physics, Universität Hamburg, Luruper Chaussee 149,
      22761 Hamburg, Germany}}%
\newcommand{\uct}{\affiliation{University of Chemistry and Technology, Prague, Technická 5, 166 28
      Prague 6 -- Dejvice, the Czech Republic}}%
\newcommand{\vsb}{\affiliation{IT4Innovations, VSB -- Technical University of Ostrava, 17. listopadu
      2172/15, 708 33 Ostrava, Czech Republic}}
\newcommand{\jkemail}{\email[]{jochen.kuepper@cfel.de}}%
\newcommand{\psemail}{\email[]{petr.slavicek@vscht.cz}}%
\newcommand{\ivemail}{\email[]{ivo.vinklarek@cfel.de}}%
\newcommand{\cmiweb}{\homepage[URL: ]{https://www.controlled-molecule-imaging.org}}%
\newcommand{\photoxweb}{\homepage[URL: ]{https://photox.vscht.cz/}}%

\newcommand{\emg}{\ensuremath{\text{EMG}}\xspace}
\newcommand{\Dzero}{\ensuremath{\text{D}_0}\xspace}
\newcommand{\Done}{\ensuremath{\text{D}_1}\xspace}
\newcommand{\Dtwo}{\ensuremath{\text{D}_2}\xspace}
\newcommand{\Dthree}{\ensuremath{\text{D}_3}\xspace}

\begin{document}

\title{Proton transfer and hydronium formation in ionized water}
\author{Ivo S.\ Vinkl{\'a}rek}\ivemail\cfeldesy%
\author{Sebastian Trippel}\cfeldesy\uhhcui%
\author{Michal Belina}\uct\vsb%
\author{Luisa Blum}\cfeldesy\uhhphys%
\author{Hubertus~Bromberger}\cfeldesy%
\author{Petr Slav{\'i}{\v c}ek}\psemail\photoxweb\uct%
\author{Jochen Küpper}\jkemail\cmiweb\cfeldesy\uhhphys\uhhcui%
\date{\today}%
\begin{abstract}\noindent
   Aqueous radiation chemistry emerges through ultrafast proton transfer and ion-radical formation
   with unexplored energy-redistribution dynamics steering the subsequent reactions. We performed a
   time-resolved disruptive-probing experiment on pure water dimer, \HHOd, to disentangle the
   elementary post-ionization reactions. Through kinetic-energy-resolved ion imaging, we unraveled
   the dynamics in the \HHOdp ground state: at low-energy ($\ordsim$0.05~eV) ultrafast proton
   transfer ($\ordsim$19~fs) is followed by $\HHHOp+\text{OH}$ fragmentation ($\ordsim$360~fs). At
   higher energies, proton transfer becomes hindered ($\ordsim$60~fs) while the subsequent
   fragmentation becomes faster ($\ordsim$210~fs), eventually ($>0.15$~eV) merging into coupled
   dynamics ($\ordsim$100~fs). Additionally, we observed \HHOdp stabilization proceeding through a
   Zundel-like structure. These timescales and product energies reveal how ion-radical formation in
   ionized hydrogen-bonded networks shapes reactivity in aqueous dynamics.
\end{abstract}
\maketitle%





Radiation chemistry in aqueous environments is largely governed by water
radiolysis~\cite{Alizadeh:CR112:5578, Garrett:CR105:355}. Upon irradiation, water undergoes reaction
cascades initiated by the formation of highly reactive ion–radical species~\cite{
   Loh:Science367:179, Schnorr:SciAdv9:7864}, which underpin biological radiation damage, catalysis,
material degradation, and radical-driven oxidation~\cite{Garrett:CR105:355}. A molecular-level
characterization of the formation of transient reactive species and the associated energy
redistribution is essential for advancing technologies including radiation
therapy~\cite{Garrett:CR105:355, Jonah:RadiatRes144:141}, wastewater
treatment~\cite{Jiang:EPSE39:13294, Wojnarovits:JRNC311:973}, nuclear–reactor-coolant
systems~\cite{Garrett:CR105:355, Christensen:NucTech131:102}, and spaceflight
applications~\cite{Fang:Nutrition18:872, Garrett-Bakelman:Science364:eaau8650}.

Surprisingly, the initial elementary steps driving the subsequent post-ionization dynamics in water
were only recently explored in a time-resolved fashion~\cite{Loh:Science367:179,
Schnorr:SciAdv9:7864, Li:Science383:1118}. These studies provide a coarse picture lacking state- and
energy-resolved details. The dominant initial step following valence ionization of liquid water is
the ultrafast proton transfer (PT) reaction occurring on a sub-50-fs
timescales~\cite{Marsalek:JCP135:224510, Loh:Science367:179, Kamarchik:JCP132:194311}, producing a
hydronium cation (\HHHOp) and a hydroxyl radical (OH):
\begin{equation} \notag%
   \text{H}_2\text{O}\cdots{}\text{H}_2\text{O}
   \xrightarrow[\text{ionization}]{-e^-_\text{valence}} \HHOp{}\cdots\text{H}_2\text{O}
   \xrightarrow{\text{PT}} \HHHOp{}\cdots{}\text{OH}
\end{equation}
These products drive secondary PTs, oxidative reactions, or undergo recombination. Early pump-probe
studies lacked the temporal resolution to capture the initial PT dynamics
\cite{Marsalek:JCP135:224510}. Time-resolved x-ray absorption spectroscopy (XAS), enabled by x-ray
free-electron lasers, revealed an ultrafast $\sim$46~fs PT event~\cite{Loh:Science367:179}, while
ultrafast electron diffraction captured \HHHOp{}$\cdots$OH formation in $\ordsim140$~fs followed by
the ion–radical separation occurring on a $\sim$250~fs timescale~\cite{Lin:Science374:92}. These
seemingly different timescales reflect the complementary sensitivities of the probes: XAS probes
local electronic structure and diffraction probes atomic motion. Recently, experiments on
water-dimer clusters ionized by 24~eV photons reported a PT time of
$\sim$55~fs~\cite{Schnorr:SciAdv9:7864}, consistent with the earlier XAS results.

\begin{figure*} 
   \includegraphics[width=\linewidth]{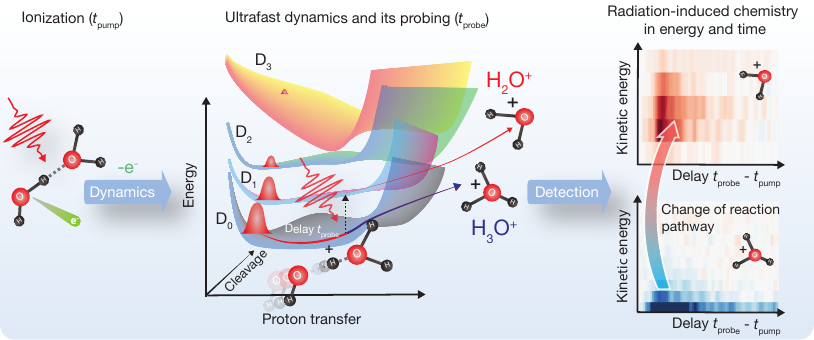}%
   \caption{\textbf{Schematic of the experiment.} Water dimers were strong-field ionized to launch
      the ultrafast dynamics. The dynamics were perturbed by a delayed, weak probe pulse that
      redistributed population and modified the fragmentation pattern. Monitoring all ion species
      over the pump–probe delay provided time-resolved signatures of PT, ion–radical stabilization,
      or fragmentation. Velocity-map imaging further yielded delay-dependent momentum distributions
      that reveal the interplay between PT and fragmentation.}
   \label{fig:Intro}
\end{figure*}

Molecular clusters permit direct detection of ionic fragments, providing detailed insights into
energy redistribution along ultrafast reaction pathways~\cite{Vinklarek:PCCP23:14340}, whereas
secondary processes characteristic of the bulk liquid~\cite{Goodwin:CR124:7379} remain inaccessible.
We used water dimer, \HHOd, a controllable and well-defined model~\cite{Vinklarek:JPCA128:1593,
Bieker:JPCA123:7486}, to investigate post-ionization chemistry in aqueous systems. Although \HHOd
does not reproduce the extended hydrogen-bond network of liquid water, it provides direct access to
the earliest elementary step of ionization-driven processes within a hydrogen-bonded water pair. The
recent controversy regarding the role of spontaneously formed $\HHHOp\cdots\text{OH}$ and persistent
\HHOdp species in water aerosols~\cite{Xing:AC134:e202207587, Chen:ACIE63:e202407433} further
underscores the importance of understanding the smallest ionized water aggregates. Moreover, the
absence of solvent caging allows to directly observe the initial translation energies of the nascent
$\mathrm{OH}$ radical and \HHHOp ion, providing molecular-level constraints on the initial
conditions for subsequent caging, diffusion, recombination, and solvent-mediated chemistry in liquid
water~\cite{Cordsmeier:JACS148:8567}.

Previously, we found a surprising diversity of ion products after single ionization of
\HHOd~\cite{Vinklarek:JPCA128:1593}. Our approach using the electrostatic
deflector~\cite{Chang:IRPC34:557} allows the assignment of the originally ionized system without the
need for simultaneous detection of multiple ions, as in coincidence-detection
methods~\cite{Schnorr:SciAdv9:7864}. Similar to earlier studies~\cite{Loh:Science367:179,
   Li:Science383:1118}, we populated \HHOdp \emph{via} strong-field ionization. Our recent study
showed that this ionization predominantly accesses the four energetically lowest states,
\Dzero\ldots\Dthree~\cite{Vinklarek:water2-moadk:inprep}. According to our previous \emph{ab initio}
simulations~\cite{Schnorr:SciAdv9:7864} these undergo PT, forming the ion–radical complex
($\HHHO\cdots\text{OH})^+$, which either stabilizes, predominantly for ionization into \Dzero, or
fragments to $\HHHOp+\text{OH}$.

To explore the elementary dynamics of ionized water dimer, we employed the mass-spectrometric
\emph{disruptive-probing} (DP) pump--probe method. DP was recently presented as a panoramic
approach~\cite{Jochim:RSI93:033003,Anishchik:PCCP28:6711} that synchronously monitors
delay-dependent ion-signal changes arising from perturbations of evolving electronic and nuclear
dynamics in multiple independent photochemical pathways~\cite{Sussman:Science314:278}. A schematic
overview of the approach is shown in Figure~1: A strong-field
($I_{\text{pump}}=2.3\cdot10^{14}~\Wpcmcm$, $800$~nm) pump pulse initiates the dynamics, while a
weaker, non-ionizing NIR-probe pulse perturbs the ongoing evolution as a function of time. Assuming
that the modulation reflects the intermediate-state population, the transient depletion signal can
be interpreted as a measure of the underlying reaction-pathway dynamics, even when the disruption
mechanism is not fully specified~\cite{Jochim:RSI93:033003, Sussman:Science314:278}. A further
advantage is that these dynamics can be directly correlated with the mass-resolved product species,
linking the transient response to the observed specific reaction products. A key methodological
advance of our work is the combination of DP with velocity-map imaging, which adds kinetic-energy
resolution to the transient ion signals. This allows for every detected product species to correlate
the energy and the timescale of its formation. Complementing these experiments with \emph{ab initio}
simulations of the dynamics to 1~ps allowed us to assign fragmentation pathways beyond the initial
PT event.

We disentangled the ultrafast dynamics of \HHOdp selectively in its electronic ground state,
determined energy-resolved timescales for proton transfer ($20\ldots100$~fs), fragmentation
($360\ldots100$~fs), and their coalescence at higher energies -- as well as \HHOdp stabilization
($\ordsim1$~ps) at lower energies. Our findings provide direct insight into the energy
redistribution during hydronium \HHHOp, OH-radical, and ion--radical formation in ionized aqueous
systems and thus constrain the initial conditions for subsequent caging, diffusion, and
recombination dynamics.

\section*{Results and discussion}
\label{sec:PP-signal}
\begin{figure*} 
   \includegraphics[width=\linewidth]{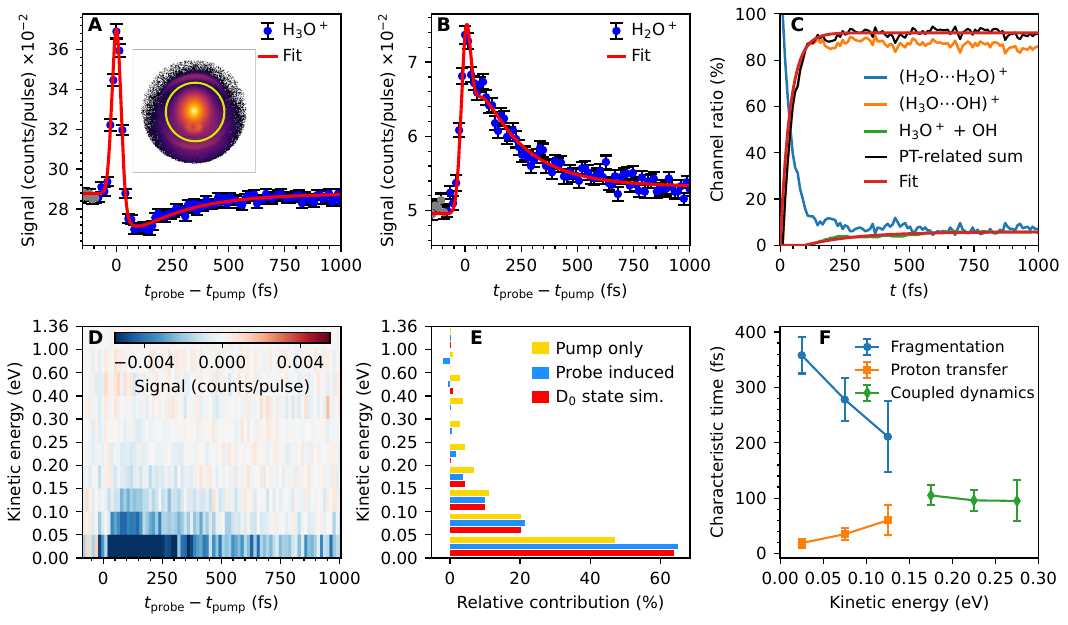}%
   \caption{\textbf{Recorded transient signals related to disruptive probing of the \HHHOp channel
      and their analysis.} \textbf{A} and \textbf{B} display transient signals of the \HHHOp and
      \HHOp channels (blue dots and black error bars), respectively. Fits (red lines) were obtained
      using the parametric model described in Materials and Methods SI1. The data (gray points) in the range
      $(-500,-50)$~fs were excluded from the fitting procedure due to probe-induced pre-excitation
      effects. %
      The inset in \textbf{A} shows a velocity map of \HHHOp with only the central part (within
      yellow circle) originating from \HHOdp dynamics. %
      \textbf{C} displays the result of simulations of \HHOdp dynamics in its ground electronic
      state (\Dzero), corresponding data for \Done{}\ldots\Dthree are provided in Supplementary Text SI2. %
      \textbf{D} represents the background corrected delay- and energy-resolved \HHHOp signal. %
      \textbf{E} shows the comparison of the bleached (blue), pump-only (yellow), and simulated
      (\Dzero, red) kinetic energy distributions~\cite{Vinklarek:water2-moadk:inprep}. %
      \textbf{F} displays the timescales of the \HHHOp formation obtained from the kinetic
      energy-resolved analysis using the parametric models described in~Materials and Methods SI1; see also
      Supplementary Text SI2 for the individual fits. }
   \label{fig:Layout-02}
\end{figure*}
\begin{figure*} %
   \includegraphics[width=0.66\linewidth]{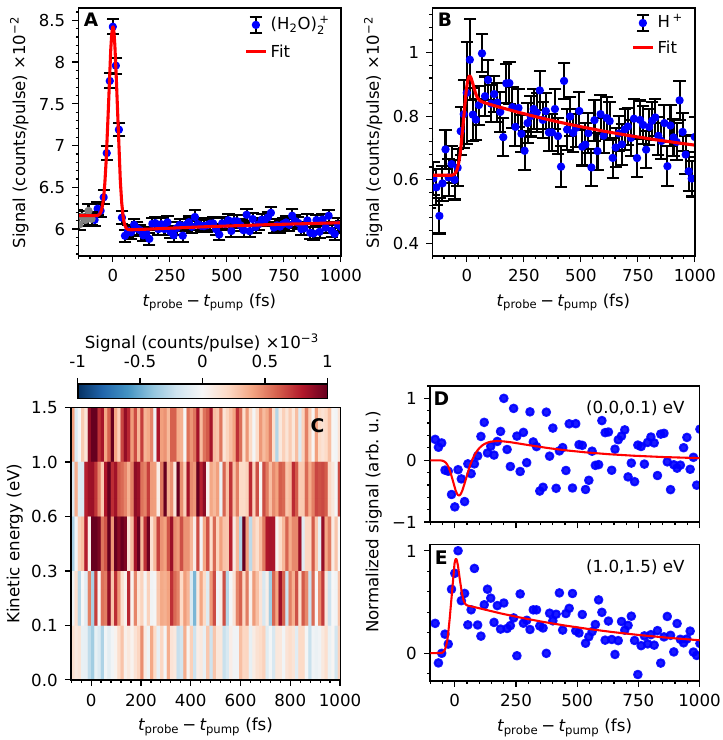}%
   \caption{\textbf{Recorded transient signals related to disruptive probing of the \HHOdp channel.}
      \textbf{A} and \textbf{B} display transient signals of \HHOdp and \Hp, respectively.
      \textbf{C} shows the delay- and energy-resolved \Hp signals after subtraction of the pump-only
      contributions. \textbf{D} and \textbf{E} show transient \Hp signals for selected kinetic
      energy ranges. All fits (red lines) were obtained using the parametric model described in
      Materials and Methods SI1. The data (gray points) in the range $(-500,-50)$~fs were excluded from the
      fitting procedure due to probe-induced pre-excitation effects.}
   \label{fig:Layout-03}
\end{figure*}
We recorded transient ion-momentum signals capturing the evolution of the \HHOdp reaction pathways,
illustrated by selected channels of \HHHOp and \HHOp in \autoref[A,~B]{fig:Layout-02} and of \HHOdp
and \Hp in \autoref[A,~B]{fig:Layout-03}. In the inset of \autoref[A]{fig:Layout-02}, we show an
exemplary velocity-map image of \HHHOp; the transient signals were extracted from the central
regions (within yellow circle) using the three-dimensional momentum distributions, thus excluding
contributions from \HHOdpp. The strong cross-correlation features peaking at time zero are due to
overlap of the pump and probe pulses and are not discussed further. All ion species display clear
time-dependent behavior at positive pump-probe delays $\Delta{t}=t_\text{probe}-t_\text{pump}>0$. In
particular, the \HHOdp and \HHHOp transient signals in \autoref[A]{fig:Layout-02} and
\autoref[A]{fig:Layout-03}, respectively, show a time-dependent bleaching, while all other detected
channels exhibit an enhancement of the signal relative to the pump-only level. The total ion yield
across all signals, outside the cross-correlation region, remains constant. The DP field perturbs
the ongoing fragmentation dynamics in \HHOdp, resulting in a redistribution of the ion signals
across different channels. This redistribution can be tracked by analyzing correlations between the
transient signals of individual ion species. Comparing the transient signals with one another
reveals that the disruption of \HHHOp formation enhances the \HHOp, \OHp, and \Op channels, while
the bleached \HHOdp signal is transferred to the \Hp channel.

All acquired pump–probe signals were fitted using a parameterized model based on a linear
combination of exponentially-modified-Gaussian (\emg) functions shown as red traces in
\autoref{fig:Layout-02} and \ref{fig:Layout-03}. Model details and extracted parameters for all
measured transient signals are given in Materials and Methods SI1 and Supplementary Text SI2. While the \HHOdp, \Hp, and \Op traces are well
described by a single-term \emg model, the other ion species, \HHHOp, \HHOp, and \OHp, require a
double-term \emg model. Three distinct timescales emerge in the acquired data: The first one
represents an ultrafast process of 25\ldots40~fs, corresponding to the drop in the \HHHOp trace. A
second timescale of several-hundred femtoseconds corresponds to the recovery of the \HHHOp signal,
\ie, 200\ldots250~fs for \HHHOp, \HHOp, and \OHp and $\ordsim580$~fs for the \Op transient. The
third timescale, in the \HHOdp and \Hp signals, is on the order of 1~ps. The identification of the
bleached \HHHOp and \HHOdp channels as the sources of the observed transients ensures that the
extracted timescales correspond directly to the $\HHHOp+\text{OH}$ fragmentation and \HHOdp
stabilization pathways. Thus, the PT and fragmentation dynamics leading to \HHHOp formation are
characterized by lifetimes of $39\pm9$~fs and $250\pm28$~fs. The \HHOdp stabilization is largely
completed at $\ordsim$1~ps.

The experimental observables were interpreted using \emph{ab initio} calculations using nonadiabatic
surface-hopping dynamics at the extended multistate complete-active-space second-order perturbation
(XMS-CASPT2) theory level~\cite{Schnorr:SciAdv9:7864}. The simulations, initialized in the
$\Dzero\ldots\Dthree$ states and extended to 1~ps, provided time-dependent branching ratios of the
dominant channels. \autoref[C]{fig:Layout-02} illustrates the total yield of
$[\HHHO{}\cdots\text{OH}]^+$ and \HHHOp products through PT and fragmentation in \Dzero.
Single-exponential fits (red traces) to the $[\HHHO{}\cdots\text{OH}]^+$ formation and \HHHOp
fragmentation channels yield the PT and fragmentation timescales. Overall, we obtained excellent
agreement of the simulated PT ($38\pm1$~fs) and fragmentation ($259\pm19$~fs) timescales with the
corresponding experimental timescales derived from the total \HHHOp signal. Note that the
simulations include nuclear quantum effects through the initial-condition sampling, while the
subsequent trajectories are propagated with classical nuclei. The calculated longer-time
mode-specific relaxation should be interpreted qualitatively, see Materials and Methods SI1. In
contrast, for the simulated dynamics starting in the excited states $\Done\ldots\Dthree$, the
timescales for PT and fragmentation differ markedly from the experimental ones: For PT the
simulations yield $70\ldots85$~fs and for fragmentation they are either considerably slower (\Done
\& \Dtwo, $\ordsim400$ \& $\ordsim725$~fs) or faster (\Dthree, $\ordsim200$~fs). This is an
intriguing finding, as it suggests the selectivity of the DP technique for our experimental
conditions. The disruption itself may proceed \emph{via} a one- or few-photon transition to a
different potential energy surface or through a dynamic Stark shift of the involved surface(s), \eg,
barrier suppression~\cite{Jochim:RSI93:033003}. We assume that the observed selectivity for \Dzero
arises from a resonant one-photon transition between the \Dzero and \Done states, whose
ionization-energy difference closely matches the single-photon energy at 800~nm ($\ordsim 1.5$~eV),
as indicated in \autoref[middle panel]{fig:Intro}.

\subsection*{Energy-resolved chemistry}
\label{sec:Hydronium}
A kinetic-energy-resolved analysis of the \HHHOp-formation dynamics yields continuously varying
timescales for PT and fragmentation dynamics. The \HHHOp data were divided in ten kinetic-energy
ranges from 0 to 1.36~eV, see \autoref[D]{fig:Layout-02}, to which we individually fitted the
parametric \emg model. These fits were used for background correction of the recorded
energy-resolved data. The resulting delay- and kinetic-energy–resolved \HHHOp signals are shown in
\autoref[D]{fig:Layout-02}. A negative signal corresponds to the probe-induced transfer of \HHHOp
population to other fragmentation channels. Analogous to a bleach signal in conventional transient
absorption spectroscopy~\cite{Vinklarek:CP529:110542}, this transient depletion is tracked as a
function of delay to extract the energy-dependent dynamics of PT and fragmentation leading to
\HHHOp. The \HHHOp kinetic-energy redistribution induced by the probe is obtained through
integration of these signals along the pump-probe-delay axis. It shows a vast contribution at
kinetic energies $\smaller0.3$~eV, see \autoref[E]{fig:Layout-02} (blue). This probe-induced
kinetic-energy distribution (KED) is clearly different from the observed pump-only \HHHOp KED
(yellow) and in excellent agreement with the \HHHOp KED obtained from simulations for ionization to
the \Dzero state of \HHOdp (red)~\cite{Vinklarek:water2-moadk:inprep}. This supports our assignment
of observing chemical dynamics in the \Dzero state, \emph{vide supra}.


\begin{figure*} 
   \includegraphics[width=0.67\linewidth]{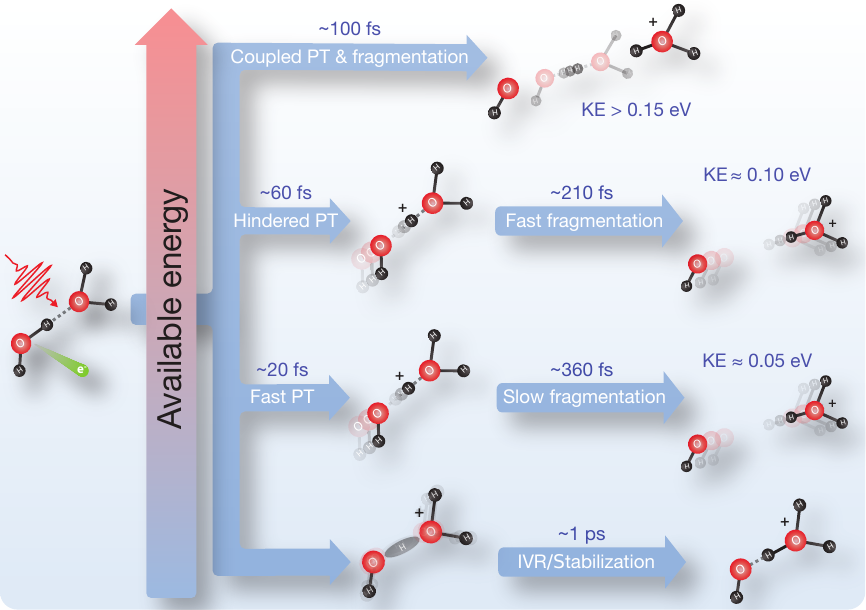}%
   \caption{\textbf{The observed \HHOdp{} dynamics.} Energy-resolved PT and fragmentation pathways
      are identified from the \HHHOp signal and low-energy stabilization \emph{via}
      inter-to-intra-molecular vibrational relaxation (IVR) is observed as the \HHOdp signal.}
   \label{fig:Summary}
\end{figure*}

An inspection of \autoref[D]{fig:Layout-02} reveals distinct timescales associated with the
individual kinetic-energy ranges. The timescales extracted for $0\ldots0.3$~eV are summarized
in~\autoref[F]{fig:Layout-02}. This shows an increase of the fast-component timescale from 19~fs to
60~fs, associated with PT (orange), with a \HHHOp kinetic-energy increase from 0~eV up to 0.15~eV.
This slowing demonstrates the transition from a direct to a hindered PT process, indicating changes
in the energy flow on the potential energy landscape. This is accompanied by a simultaneous decrease
of the slow-component, fragmentation, timescale from 358~fs to 211~fs, associated with faster
fragmentation at higher energies (blue). At higher kinetic energies, between 0.15~eV and 0.3~eV, the
two timescales converge to $\ordsim100$~fs, obtained from a modified model assuming equal PT and
fragmentation timescales. Thus, strikingly, the \HHHOp formation dynamics exhibits a strong
kinetic-energy dependence, with its characteristic timescale changing markedly across the
released-energy range, see \autoref{fig:Summary}.

Our kinetic-energy–resolved analysis provides insight beyond an electronic-state-averaged
description. Interestingly, the higher-energy converged-single-timescale \HHHOp coincides with the
110~fs delocalized-proton lifetime in aqueous hydroxide solution~\cite{Roberts:PNAS106:15154}, and
it is comparable to the computed period ($\ordsim380$~\invcm/$\ordsim90$~fs) of the O$\cdots$O
stretch vibrational mode of \HHOdp \cite{Talbot:JACS138:11936}. The shortening of the
$\text{O}\cdots\text{O}$ distance was identified as a key factor governing PT
dynamics~\cite{Svoboda:PCCP15:11531}. We hypothesize that the observed coupled PT and fragmentation
dynamics for higher kinetic energies arises from O-O-stretch excitation, leading to vibrationally
driven direct dissociation and thus enhanced kinetic-energy
release~\cite{Grygoryeva:aipadv9:0351511}. Such control \emph{via} intermolecular vibrational modes
may allow steering the energy carried by hydronium and the OH radical, shaping their roles in
radiation chemistry.

\subsection*{Bound-state dynamics} 
\label{sec:Stabilization}
We also observed a signal bleach for the \HHOdp transient signal shown in
\autoref[A]{fig:Layout-03}, reflecting the stabilization dynamics of this parent ion. \HHOdp does
not have kinetic energy from fragmentation, preventing a kinetic-energy–resolved analysis.
Nevertheless, the DP induced the transfer of \HHOdp population into the \Hp channel,
\autoref[B]{fig:Layout-03}. These \Hp signals were analyzed in an energy-resolved manner analogous
to the \HHHOp case, providing the energy-resolved signals of \Hp shown in
\autoref[C]{fig:Layout-03}. The time-dependent signals for the energy ranges $0\ldots0.1$~eV and
$1.0\ldots1.5$~eV are shown in \autoref[E]{fig:Layout-03}. The \Hp signal in the $0\ldots0.1$~eV
range shows a rapid signal bleach in the first 100~fs, the origin of which remains unresolved. Most
\Hp appear with kinetic energies of 0.3\ldots1.5~eV, with a timescale of $0.7$\ldots$1.4$~ps.

We ascribe the \Hp signal to the disruption of transient, non-stabilized \HHOdp and thus the
observed picosecond timescale to \HHOdp stabilization, as illustrated in \autoref{fig:Summary}.
However, in isolated \HHO molecules and OH radicals the covalent O-H bonds have dissociation
energies of about 5.1~eV and 4.4~eV, respectively~\cite{Maksyutenko:JCP125:181101,
   Ruscic:JPCA106:2727}; the O-H bonds in isolated \HHHOp are even stronger~\cite{Bodi:CS5:3057}.
Cleaving such bonds would require the absorption of three to four 800~nm photons. Even though the
hydronium-ion-hydroxyl-radical pair, like Zundel complexes, has a strong proton delocalization along
its O-O axis~\cite{Vendrell:JCP127:184303}, the proton is overall still strongly bound.
Consequently, our weak probing pulse cannot drive the stabilized ion-radical \HHOdp pair into the
proton channel.

We assume that the weak probe pulse can eject the proton only during the early stage of the
dynamics, before the reaction energy is redistributed from intermolecular modes, \eg,
proton-shuttling and O-O stretch, to intramolecular modes to form a stabilized \HHOdp. In other
words, the dynamics observed on the hundreds-of-femtoseconds to picosecond timescale are attributed
to the decay of a transient Zundel-like \HHOdp, ultimately leading to the formation of a stabilized
ion–radical pair $\HHHOp\cdots\text{OH}$. In the liquid, Zundel-like structures decay substantially
faster (100–200~fs)~\cite{Woutersen:PRL96:138305, Roberts:PNAS106:15154}. Nevertheless, these
sub–200 fs components mainly represent localized proton fluctuations caused by short-lived
structural rearrangements, rather than a stable, fully formed Zundel configuration, which also in
the bulk persists for much longer times ($\larger480$~fs)~\cite{Thamer:Science350:78}. Analyzing in
how far this relates to possibly spontaneously formed \HHOdp species in water
aerosols~\cite{Xing:AC134:e202207587, Chen:ACIE63:e202407433} is beyond the scope of our manuscript.

A remaining open question concerns the electronic-state contribution to the observed stabilization
dynamics of \HHOdp. Based on Ammosov-Delone-Krainov tunneling-ionization
calculations~\cite{Vinklarek:water2-moadk:inprep}, the majority of the \HHOdp signal detected in the
pump-only experiment~\cite{Vinklarek:JPCA128:1593} must stem from initially populated \Dzero and
\Done states. Unlike the assignment of the \HHHOp dynamics to the \Dzero state from comparison to
our computational modeling (\emph{vide supra}), the \HHOdp signals cannot be uniquely attributed to
a single electronic state. Considering the equivalent of Kasha's rule that electronically excited
ions undergo very rapid internal conversion to their electronic ground state, resulting in
significantly vibrationally excited \HHOdp in their D$_0$ state that would then fragment, we assume
that all stabilized \HHOdp is formed following ionization to the D$_0$ state.

\bigskip

Overall, we unraveled the initial elementary processes after ionization of water,
exploiting the \HHOdp system, with a focus on the appearance of the primary ionic products,
including hydronium ion (\HHHOp) and stabilized \HHOdp. Utilizing the energy distributions of the
reaction products, our results go beyond electronic-state–averaged descriptions. As illustrated in
\autoref{fig:Summary}, low-energy stabilization of \HHOdp on picosecond timescales proceeds through
Zundel-like configurations that stabilize by atomic rearrangement and vibrational relaxation from
intermolecular to intramolecular modes. For increasing energies, we see an energy-driven evolution
from sequential ultrafast proton transfer with subsequent fragmentation of the ion-radical pair to
an intermediate-timescale concerted pathway. The measured kinetic-energy distribution of the
$\text{OH}+\HHHOp$ channel provides a molecular-level benchmark for models of the initial caging,
energy dissipation, recombination, and diffusion of nascent ion–radical pairs in liquid
water~\cite{Cordsmeier:JACS148:8567}.

Our results demonstrate that combining species-controlled samples, multi-mass velocity-map imaging,
and disruptive probing with computational-dynamics simulations allows us to disentangle the
ultrafast dynamics of complex molecular systems in energy and time. The agreement of our
energy-averaged data with earlier ultrafast post-ionization dynamics observed in
clusters~\cite{Schnorr:SciAdv9:7864} and liquids~\cite{Loh:Science367:179} demonstrates that our
observations are directly relevant to initial-stage liquid-phase chemistry: Our results show that
early ion--radical chemistry in water is governed by energy redistribution, not a single prompt
proton-transfer step.

The sensitivity to internal dynamics, such as the inferred Zundel-like structures, provides
opportunities for the investigation of internal motions including general isomerizations. Our
approach further opens important prospects for probing biomolecules in their native aqueous
environment~\cite{Onvlee:NatComm13:7462}. Coincidence measurements combining the ion imaging with
electron spectroscopy offer direct access to the coupled dynamics of nuclear and electronic reaction
pathways. This provides a basis for future attempts to tailor wavepackets to steer ion–radical
chemistry in aqueous environments.

\bibliography{string,cmi}%

\section*{Acknowledgments}
We gratefully acknowledge helpful discussions with Denis S. Tikhonov and Emil Vogt.

\paragraph*{Funding:}
This work was supported by Deutsches Elektronen-Synchrotron (DESY), a member of the Helmholtz
Association (HGF), and by the Cluster of Excellence ``Advanced Imaging of Matter'' of the Deutsche
Forschungsgemeinschaft (DFG, AIM, EXC~2056, ID~390715994). Data processing and analysis were carried
out using the Maxwell computing resources at DESY. The endstation for controlled-molecule
experiments (eCOMO) was funded with support from the Center for Molecular Water Science. I.S.V.\
acknowledges support from the Alexander von Humboldt Foundation. M.B. and P.S. thank to Czech
Science foundation (projects numbers 23-07066S and 26-22810S).

\paragraph*{Author contributions:}
Original concept: I.S.V., J.K.;
Performed experiments: I.S.V., L.B.;
Analysis of data: I.S.V., L.B.;
Trajectory based simulations: M.B., P.S.;
Software and experimental support: H.B.;
Original draft: I.S.V.;
Discussion: I.S.V., S.T., M.B., P.S., J.K.;
Writing and editing: I.S.V., S.T., P.S., J.K.;

\paragraph*{Competing interests:}
There are no competing interests to declare.

\paragraph*{Data and materials availability:}
All data underlying the figures and all analysis scripts are provided as Supplementary Material. All
processed data are publicly available at Zenodo (DOI: 10.5281/zenodo.18678891,
\href{https://zenodo.org/records/18678891?preview=1&token=eyJhbGciOiJIUzUxMiJ9.eyJpZCI6ImRiM2Q3M2ZkLTJjNTgtNDAwYy04ZjVjLTRlOTQ2YWUzMzJmZCIsImRhdGEiOnt9LCJyYW5kb20iOiIyODExYjllZDI5MGMxMzQ0NDcwODMzZjdlM2RkNjQ2MCJ9.M2Ac7sIAikqEz4dnckar-ZDGWmkWKEJzwSTYpsbDAx52m0j_RAsq5UwyQMgdOn2mBPZUP7qDl01ag-lGjE-CFw}{preview
   link}). All raw data are archived in DESY's data repositories under ID~11019141 and are available
upon reasonable request.

\subsection*{Supplementary materials}
\noindent%
Materials and Methods SI1\\
Supplementary Text SI2\\
Tutorial on disruptive probing SI3\\
Figures SI1 to SI31\\
Tables SI1 to SI4\\
References \emph{(1-27)}\\

\onecolumngrid
\clearpage
\listofnotes%
\end{document}


\title{Supplementary information: Proton transfer and hydronium formation in ionized water}%
\author{Ivo S.\ Vinkl{\'a}rek}\ivemail\cfeldesy%
\author{Sebastian Trippel}\cfeldesy\uhhcui%
\author{Michal Belina}\uct\vsb%
\author{Luisa Blum}\cfeldesy\uhhphys%
\author{Hubertus Bromberger}\cfeldesy%
\author{Petr Slav{\'i}{\v c}ek}\psemail\photoxweb\uct%
\author{Jochen Küpper}\jkemail\cmiweb\cfeldesy\uhhphys\uhhcui%
\date{\today}%
\maketitle%

\subsubsection*{This PDF file includes:}
\noindent
\hyperref[sec:Methods]{Materials and Methods}\\
\hyperref[sec:extraresults]{Supplementary Text}\\
\hyperref[sec:tutorial]{Tutorial on disruptive probing}\\
Figures S1 to S31\\
Tables S1 to S4\\
References \emph{(1-27)}\\

\clearpage

\section{Materials and Methods}
\label{sec:Methods}
\subsection{Experimental methods and data analysis}
\label{sec:Exp-methods}
\subsubsection{Preparation of water dimer ensemble}
\label{sec:HHO-ensemble-method} All the experiments were performed using the endstation for
controlled molecule (eCOMO) experiments~\cite{Jin:RSI96:023305}. A molecular beam containing water
clusters was generated by supersonic expansion of water mixture with helium buffer gas
($p_{\text{He}}= 100~\text{bar}$) using an Even-Lavie valve~\cite{Even:EPJTI2:17} running at 100~Hz
repetition rate and stabilized at a temperature of 50~$^{\circ}$C. The expanded gas was then skimmed
by the first skimmer ($= 3$~mm) into the deflector chamber, where it was further collimated by a
second skimmer ($= 1.5$~mm) before entering the electrostatic deflector. The inhomogeneous
electrostatic field ($E_\text{d} \ordsim 100$~kV/cm, $U_{\text{d}} = 22$~kV) resulted in spatial dispersion of the various water
clusters according to their effective dipole-moment-to-mass ratios~\cite{Chang:IRPC34:557}. Behind
the deflector, the molecular beam was further cut by the knife edge to improve separation of the
various water clusters~\cite{Trippel:RSI89:096110}. The estimated purity of the water dimer beam was
around 96\%. Finally, the molecular beam was introduced to a VMI chamber by the third skimmer
($= 1.5$~mm) and was crossed with the IR laser radiation at the interaction center of the VMI
spectrometer with a perpendicular geometry towards the molecular beam velocity
direction~\cite{Eppink:RSI68:3477}. Interaction of the IR laser with the deflected water dimer
particles caused their ionization, which then enabled the ionized particles extraction towards the
detector. The ion products were imaged by the combination of microchannel plates in Z-stack
($U_{\text{fMCP}}=-2.4$~kV) and phosphor screen (P47, $U_{\text{PS}}= 4.3$~kV), which scintillated
the ion signal into the flashes recorded by the ultrafast Timepix3 camera with a temporal resolution
of 1.7~ns~\cite{Bromberger:JPB55:144001}. The precise ion signal time-of-flight recording enabled us
to reconstruct the 3D momentum Newton sphere for each detected ion signal. The data were processed
using PymePix software~\cite{AlRefaie:JINST14:P10003, pymepix:gitlab}. To resolve the physical
events, we used a clustering and centroiding algorithm to retrieve the time of flight and impact
position of each ion.

\subsubsection{Disruptive probing}
\label{sec:Disruptive-probing-method} The disruptive probing pump--probe scheme was applied to study
the cluster fragmentation after its single ionization~\cite{Jochim:RSI93:033003}. First a strong
ionizing pulse at 800~nm (peak intensity $I_{\text{pump}} = 2.3\cdot10^{14}~\Wpcmcm$, temporal width
$\sigma^\text{pump} = 18.9$~fs, and Keldysh parameter $\gamma_\text{K} = 0.65$, linear polarization)
triggered the dynamics followed by a second weaker probing pulse at 800~nm ($I_{\text{probe}} = 0.4
\cdot10^{14}~\Wpcmcm$, $\sigma^\text{probe}_{\text{FWHM}} = 18.9$~fs) with a temporal delay $\tau$.
The pulses were cross-polarized, with the pump pulse p-polarized (parallel with the detector plane)
and the probe pulse s-polarized. The intensity of the second probing pulse was set such that no ions
were observed from the probe-only data acquisition. The purpose of the second pulse was to cause
disruption of the ongoing water dimer dynamics, resulting in a change in the resulting ion yields,
which were then measured as a function of the pump--probe delay.

\subsubsection{Scheme of disruptive probing setup}
\label{sec:DP-table}

\begin{figure*}
   \includegraphics[width=\linewidth]{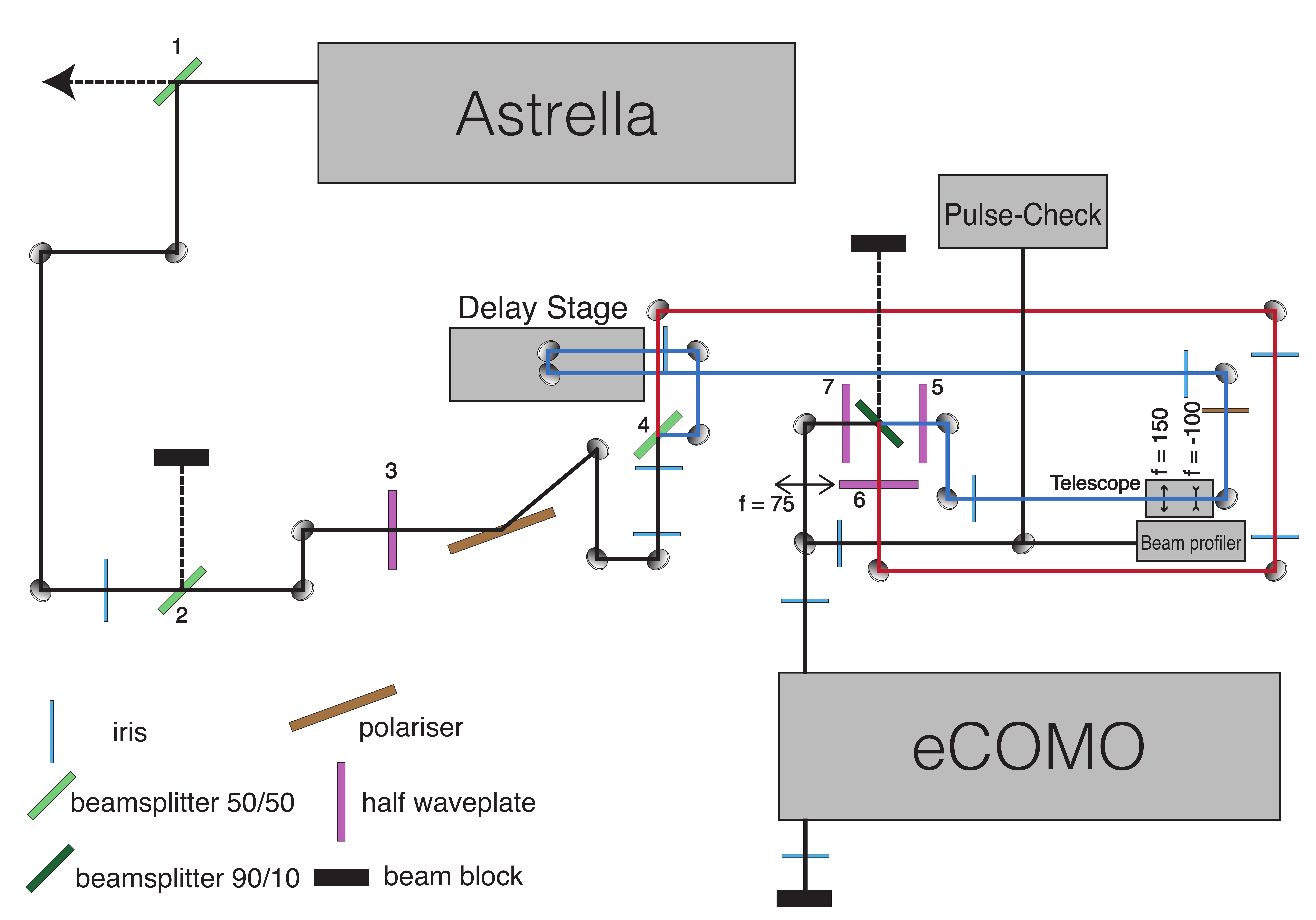}%
   \caption{Schematic picture of the pump--probe setup used for the disruptive probing experiment.}
   \label{fig:laser-table}
\end{figure*}

In~\autoref{fig:laser-table}, there is shown a schematic picture of the pump--probe setup for the
disruptive probing technique.
 The Astrella laser (Coherent) seeded with the Vitara laser (Coherent) and pumped by the
Nd:YAG laser provides 45~fs laser pulses at 800~nm, which intensity is controlled by a combination
of a half-waveplate and a thin film polarizer. Next, the beam is split into two paths, \ie, pump
beam path, which leads directly to the endstation for Controlled Molecule experiments (eCOMO), and
probe beam path, which leads through the delay stage and telescope. Both beams are then coupled
again on the second beamsplitter and focused by 75~mm lens into the experimental chamber. The
telescope is used to compensate for any optical errors in aligning the pump and probe focus
positions.

\subsubsection{Parametric modeling of acquired time-resolved signal using disruptive probing}
\label{sec:PP-model} The response of the targeted ion to the probing field is observed via the
time-dependent change in ion yields of individual ion species. This was shortly discussed
in~\cite{Jochim:RSI93:033003}, and we provide a broadened tutorial of the empirical description of
the effects caused by disruptive probing in~\autoref{sec:tutorial}.

To investigate the original dynamics, we apply the generalized model for fitting the pump--probe
signals build from: (1) A Gaussian $g$, representing the cross-correlation ion signal. (2) An
exponentially modified Gaussian function (\emgf), representing the sought ionic dynamics broadened
by the instrumental response function (IRF). (3) An error-function step $\theta$, describing either
signal originating in the probe-induced secondary fragmentation of the relaxed ions, or signal
originating in probe-only ionization of pump-pre-excited water dimer. (4) A constant
$[\text{Ch}^+]_0$, representing the pump-only signal level.

For each specific ion channel Ch$^+$, we model the obtained transient signal as
\begin{equation}
      \label{eq:PP-model}
      \begin{aligned} \relax[\text{Ch}^+] & \left(t; a_{\text{IRF}}, \sigma_{\text{IRF}}, \{a_k\},
\{\tau_k\}, a_{\theta}, [\text{Ch}^+]_0 \right) \\ = & g\left(t; a_{\text{IRF}}, \sigma_{\text{IRF}}
\right) + \sum_{k} \emgf_k\left(t; a_k, \tau_k, \sigma_{\text{IRF}} \right) \\ & + \theta\left(t;
a_{\theta}, \sigma_{\text{IRF}} \right) + [\text{Ch}^+]_0,
      \end{aligned}
\end{equation} where $a_{\text{IRF} / \theta}$ and $\sigma_{\text{IRF}}$ denote the amplitudes and
the width of the applied functions, respectively, and the individual elements in the sets $\{a_k\}$
and $\{\tau_k\}$ express the amplitudes and characteristic times describing the ion dynamics. Notice
that all the fitted parameters are channel-specific. We added an additional constraint of an upper
limit for the $\sigma_{\text{IRF}}$ equal to $\sigma^{\text{max}}_\text{IRF} = 20.7$, which we
obtained from fitting the cross-correlation signal of all detected ions, see
also~\autoref{sec:SI-to-sigma-t0}.

Finally, some of the signal dependencies exhibit a double-exponential character, which is
characteristic of processes involving the initial population followed by the decay of a metastable
state. In such cases, we impose an additional constraint on the amplitudes $a_{k}$ and decay times
$\tau_{k}$ in the form
\begin{equation} \frac{a_1}{\tau_1} = -\frac{a_2}{\tau_2}.
\end{equation} This condition is derived from an assumption that the final ion product is populated
from a metastable state. A detailed derivation of this constraint is provided
in~\autoref{sec:PP-Fitting}.

In the special case of converging lifetimes, \ie, when $\tau_1 - \tau_2 \to 0$, fitting the data
with a two-component \emgf model becomes unfeasible due to ill-defined amplitudes. In this limit,
the second term in~\autoref{eq:PP-model} is replaced by the convolution of the normalized Gaussian
excitation function $g$ with a causal gamma-like response function
\begin{equation}
    \label{eq:PP-model-special}
    \begin{aligned} a \cdot \left[ g(t'; \sigma_{\text{IRF}}) * \left( \frac{t'}{\tau} \,
e^{-t'/\tau} \cdot \Theta(t') \right) \right](t).
    \end{aligned}
\end{equation}

\subsubsection{Estimation of time zero and laser pulse duration}
\label{sec:SI-to-sigma-t0}

\begin{figure}
   \includegraphics[width=1.0\linewidth]{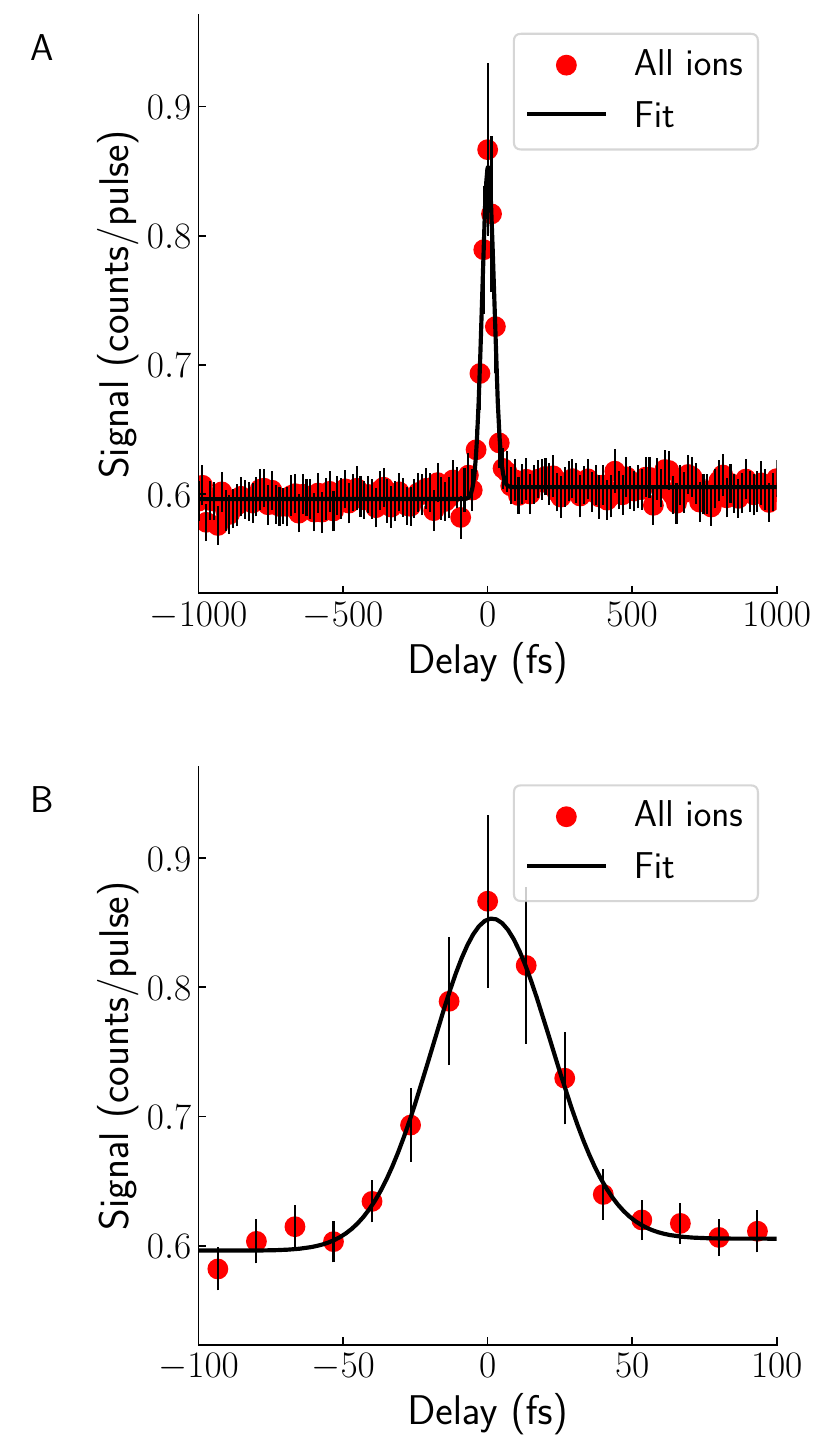}%
   \caption{(A) Fitting of $t_0$ and $\sigma_{\text{IRF}}$ from the integrated all ion signal
resulting in 1.36(84)~fs and 21.2(8)~fs, respectively. (B) Zoomed-in plot at (-100,100)~fs range.}
   \label{fig:PP-all}
\end{figure}

In the case of fitting the individual ion channel signals, we applied the same non-varying value for
the position of the zero pump--probe delay $t_0$. The value was obtained from fitting a Gaussian
function and a step function to the all-ion signal after subtraction of the background signal; see
\autoref{fig:PP-all}. The obtained value of $t^{\text{All}}_0$ is equal to 1.09(99).

Secondly, we wanted to estimate upper limit of width $\sigma_{\text{IRF}}$ characterizing the
broadening of observed molecular dynamics due to durations of pump and probe pulses. The reason for
applying a maximum value for $\sigma_{\text{IRF}}$ rather than $\sigma_{\text{IRF}}$ as a
non-varying parameter comes from the fact that individual ion channels may originate from higher
excited levels of \HHOdp, and therefore their actual $\sigma_{\text{IRF}}$ values could be narrower.
On the contrary, a significant broadening of the function width over the upper limit of
$\sigma_{\text{IRF}}$ indicates underlying dynamics with a rate constants on similar timescale as
the IRF. We obtained upper limit for $\sigma_{\text{IRF}}$ from the fitting of all ion signal, equal
to $\sigma^{\text{All}}_\text{IRF} = 20.7(8)$~fs, which characterizes the broadening caused by the
pump and probe pulse durations. Independently, we estimate the value of $\sigma_{\text{IRF}}$ using
an autocorrelator (PulseCheck, APE). Assuming a Gaussian temporal shape, the FWHM of the laser pulse
is $44.5$~fs corresponding to $\sigma^{\text{AC}}_{\text{IRF}} = 18.9(1.0)$~fs. Both values are in a
reasonable agreement. As the estimated $\sigma_{\text{IRF}}$ should be applied as an upper limit in
analysis of individual ion signal, we selected to apply $\sigma^{\text{max}}_\text{IRF} =
20.7(8)$~fs for the fitting.

In perfect conditions of the disruptive probing experiment, we would expect the signal levels to be
the same on the left and right sides of the cross-correlation signal. We observe only a very small
increase in the signal at positive pump--probe delays, which is caused by probe laser ionization.
Therefore, we can be confident that this does not cause any uncertainty regarding the origin of the
measured signals for the main channels.

\subsubsection{Probe laser focus}
\label{sec:probe-focus}

\begin{figure}
   \includegraphics[width=\linewidth]{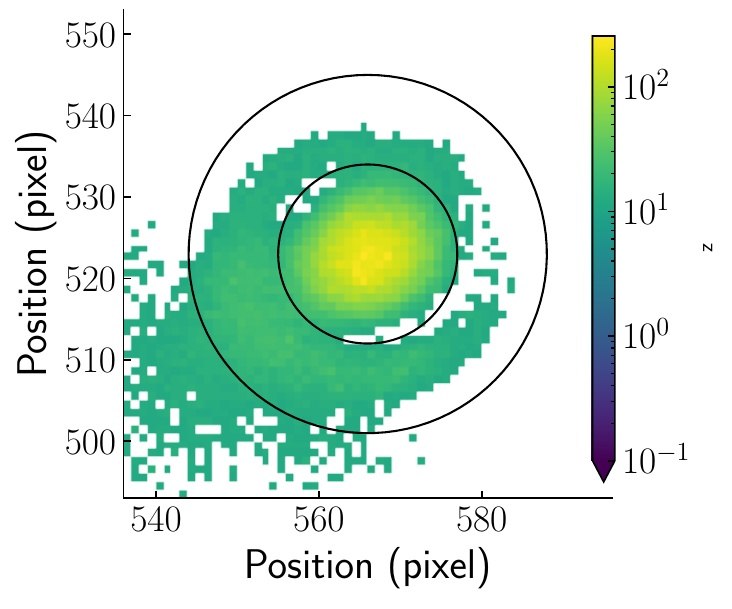}%
   \caption{Image of the probe beam laser focus taken by the laser-beam profiler. It is shown with
the logarithmic scale of the intensity axis.}
   \label{fig:probe-focus}
\end{figure}

The checking of the probe beam focus by a laser-beam profiler (CinCam CMOS-1201-Nano, CINOGY)
revealed a secondary ring structure; see the image taken and plotted in logarithmic scale
in~\autoref{fig:probe-focus}.
 This was found to be caused by diffraction on the second beamsplitter, which couples
the pump and probe beams before focusing them into the velocity map imaging (VMI) chamber.
Therefore, we integrated the signal in the ring and center areas of the focus to calculate a
correction factor for the estimated laser field intensity caused by the probing pulse. The
correction factor was estimated to be 0.7.

The focus of the pumping laser was also checked and showed a nicely Gaussian profile without any
diffraction pattern.

\subsection{Trajectory simulations}
\label{sec:Trajectory-simulations-method}
We investigated the dynamics of the water dimer following ionization into four different electronic
states, \Dzero{}–\Dthree, over a time window of up to 1~ps. These ionized states were generated by
removing an electron from the 1b$_1$- and 3a$_1$-like molecular orbitals. Compared to previous the
study~\cite{Schnorr:SciAdv9:7864} using the same method, we expanded the simulations by increasing
the number of trajectories from about one hundred to four hundred. The dynamics were treated using
the nonadiabatic surface-hopping approach in the fewest switches
formulation~\cite{Tully:JCP93:1061}. The electronic structure of the singly ionized dimer was
described using the extended multistate complete active space second-order perturbation theory
(XMS-CASPT2)~\cite{Shiozaki:WCMS8:e1331} with an active space of 11 electrons in 6 molecular
orbitals. For these calculations, the cc-pVDZ basis set was employed. All electronic structure
computations were performed with the BAGEL program, version 1.2.0~\cite{Shiozaki:WCMS8:e1331}.
Nuclear motion was propagated using a Verlet algorithm with a time step of 4 atomic units (a.u.),
$\sim$0.1~fs.

The starting conditions for the non-adiabatic molecular dynamics (NAMD) simulations were generated
from molecular dynamics (MD) employing a quantum thermostat based on the generalized Langevin
equation (GLE)~\cite{Ceriotti:PRL102:020601, Ceriotti:gle4md}. This approach incorporates nuclear
quantum effects into both, atomic positions and momenta. The subsequent nonequilibrium NAMD
trajectories were propagated with classical nuclei. A temperature of 180~K was used, together with
GLE parameters $\hbar\omega/kT = 50$ and $Ns = 50$. Thermostat matrices were downloaded from the
GLE4MD repository~\cite{Ceriotti:PRL102:020601, Ceriotti:gle4md}. The MD simulations were run with a
time step of 20~a.u. ($\sim$0.48~fs) and a total duration length of 50~ps. Energies and gradients
were obtained with the Boese-Martin for kinetics (BMK) functional~\cite{Boese:JCP121:3405} in
combination with the \mbox{6-31++g**} basis set, as implemented in the Gaussian09, revision
D.01~\cite{Frisch:G09}. Initial geometries and velocities were extracted using a fixed time
interval. The propagation of the equations of motion was carried out with the in-house ABIN package,
version 1.1~\cite{Slavicek:ABIN}.

The NAMD analysis was first based exclusively on nuclear positions. Once dissociation was observed,
the charge distribution on the fragments was determined, and if a fragment carried a charge of $+1$,
its center-of-mass kinetic energy was evaluated. Mulliken charge population analyses were performed
for every 100th geometry along each trajectory, using the Complete Active Space Self-Consistent
Field (CASSCF) method with 11 electrons in 6 active orbitals and the \mbox{6-31+g*} basis set. These
electronic structure and charge calculations were carried out with MOLPRO, version
2015.1~\cite{Werner:MOLPRO-WIREs, Werner:MOLPRO_full}.

\subsubsection{Note on treatment of nuclear quantum effects.}
Nuclear quantum effects enter the simulations through the quantum-thermostat sampling of the initial
nuclear positions and momenta, but the subsequent nonequilibrium dynamics, following ionization, are
propagated with classical nuclei. This quasiclassical treatment is expected to be most appropriate
for the early, large-amplitude proton-transfer and fragmentation motions following ionization. At
longer times, classical propagation may lead to zero-point-energy leakage, in which energy is
transferred unphysically from high-frequency O–H stretching modes into lower-frequency
intermolecular modes. Such leakage could make vibrational-energy redistribution and fragmentation
artificially efficient. Consequently, the calculated longer-time mode-specific energy redistribution
and relaxation times should be regarded as qualitative, possibly as lower-bound estimates, rather
than as exact quantum-mechanical lifetimes.

Note that the experimentally extracted timescales do not depend on this approximation, the
simulations are used to support the mechanistic and electronic-state assignments of the measured
kinetics.

\subsubsection{Parametric model for description of simulated channel dynamic}
\label{sec:Simulation-par-model}

To characterize the timescales of ion--pair \HHHOp{}$\cdots$OH formation and subsequent
fragmentation, we employed a single--exponential kinetic model~\cite{Schnorr:SciAdv9:7864}:
\begin{align}
\label{eq:Sim-model}
\begin{split} \left[ \text{P} \right](t) = a_\text{P} \cdot \left[1 - \exp\left(-\frac{t -
t_0}{\tau_{\text{P}}}\right)\right],
\end{split}
\end{align} where $a_\text{P}$ is a scaling amplitude and $\tau_{\text{P}}$ is the characteristic
time of the investigated process P. For PT, we denote the lifetime as $\tau_{\text{PT}}$, and for
fragmentation, as $\tau_{\HHHOp}$. Finally, the retardation $t_0$ arises from the methodology
applied in \emph{ab initio} simulations, where the identification of the ion--radical pair
$[\HHHO{}\cdots\text{OH}]^+$ and its subsequent fragmentation into \HHHOp is based on geometrical
constraints involving interatomic distances. The formation of the $[\HHHO{}\cdots\text{OH}]^+$ complex is
registered when the transferring proton is closer to the acceptor oxygen atom than to the donor
oxygen atom, while fragmentation is detected when the distance between \HHHOp and the OH radical
exceeds a predefined limit of 4.5 \AA.

\onecolumngrid
\clearpage
\twocolumngrid

\section{Supplementary Text}
\label{sec:extraresults}
\subsection{Overview of all detected ions}
\label{sec:Overview}
\begin{figure*}
   \includegraphics[width=\linewidth]{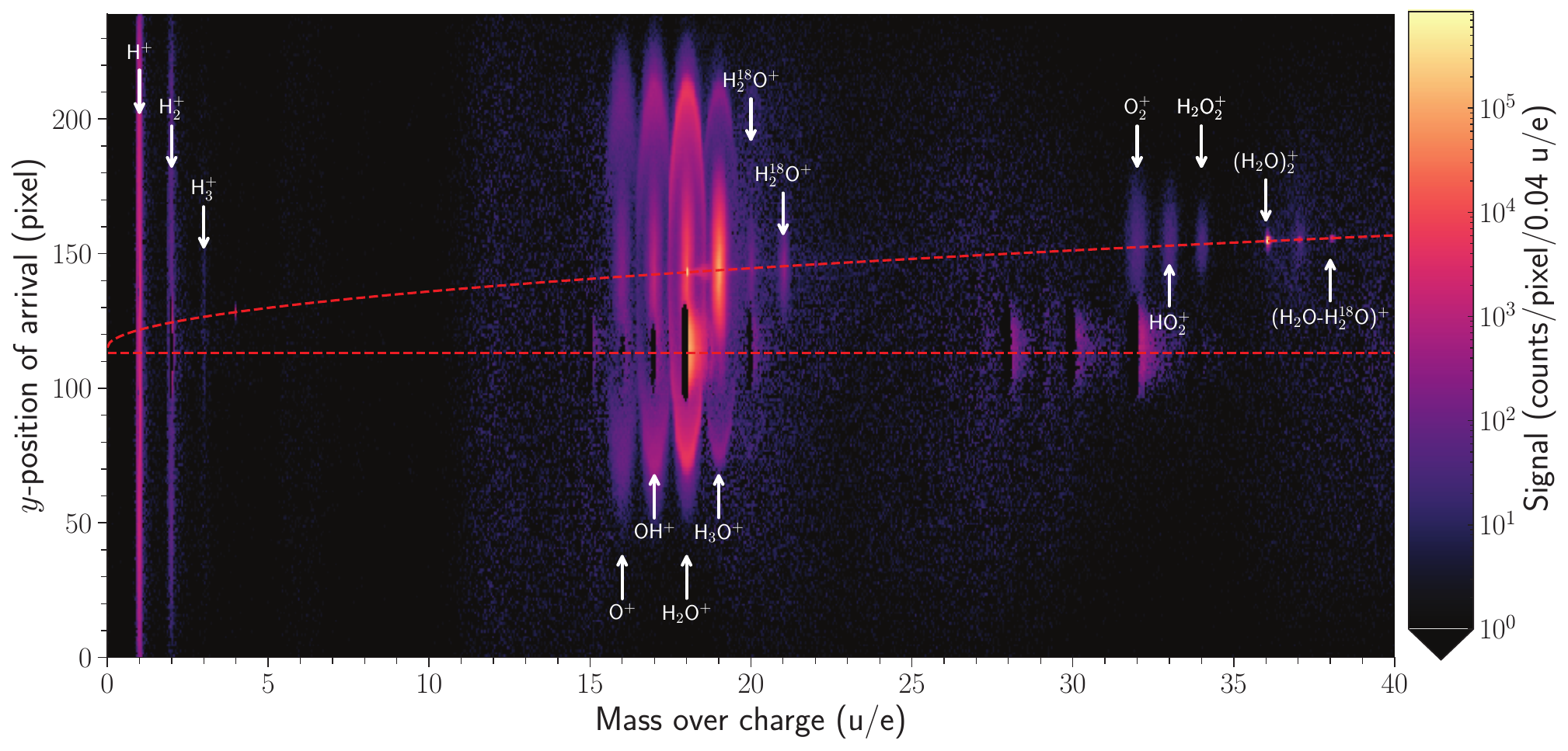}%
   \caption{Map of the ion signal according to the mass-over-charge ratio and the position of
arrival on the detector. The curved dashed line intersects the individual VMI image centers, shifted
due to the molecular beam velocity.}
   \label{fig:beamplot}
\end{figure*} We recently reported on the detected ion species and their ion yields after the
IR-laser-induced strong-field ionization of water dimer in~\cite{Vinklarek:JPCA128:1593}. The new
set of measurements at slightly different laser intensities exhibits good agreement of the detected
ion species with the previous measurements. In~\autoref{fig:beamplot}, we provide a graphic overview
of all detected ions after IR-strong-ionization of water dimers. Detailed description of the
obtained ions, different isotopologues, their ion yields, and expected appearance energies can be
found in~\cite{Vinklarek:JPCA128:1593}. Here only briefly - we observe stabilized \HHOdp, and
multiple ion fragments from \HHOdp/\HHOdpp fragmentation, \ie, H$_2$O$_2^+$, HO$_2^+$, O$_2^+$,
H$_3$O$^+$, H$_2$O$^+$, OH$^+$, O$^+$, H$_3^+$, H$_2^+$, H$^+$, and their isotopologues.

Each of the ions is represented in~\autoref{fig:beamplot} by a projection of the generated Newton
sphere after the \HHOdp/\HHOdpp fragmentation into coordinates of $y$-position of arrival and
time-of-flight recalibrated into the apparent mass-over-charge. A typical example of such a
projection is the signal of \HHOp with a broad signal blob at the center of ions populated after
\HHOdp fragmentation and a larger ring feature corresponding to ions populated after the Coulomb
explosion of \HHOdpp. At \mq{36}, we observe an intense narrow blob corresponding to the parent ion
\HHOdp~\cite{Vinklarek:JPCA128:1593}.

\subsection{Radial distributions of detected ion channels}
\label{sec:rad}

\begin{figure*}
   \includegraphics[width=0.95\linewidth]{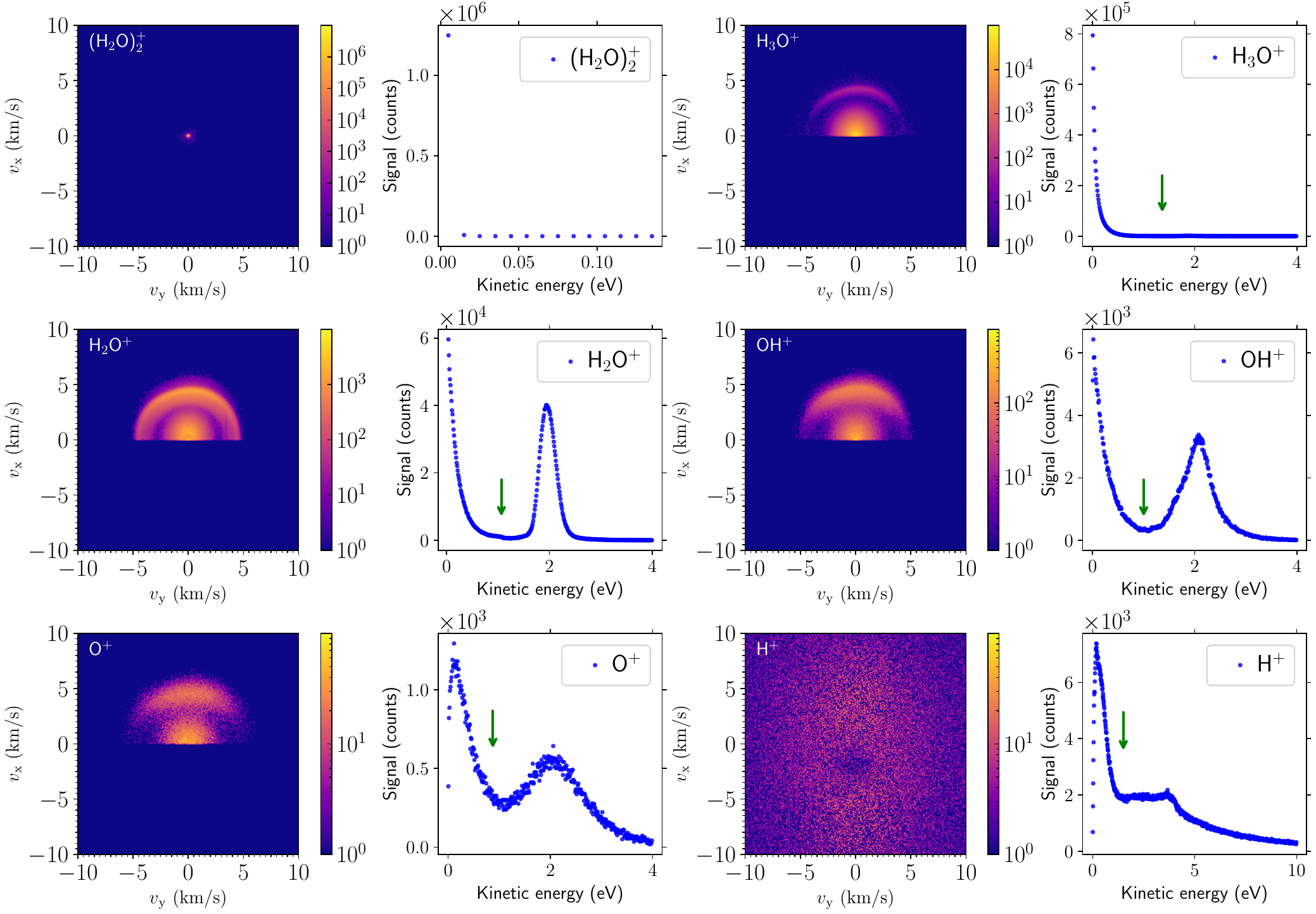}%
   \caption{Overview of ion velocity map images originating from \HHOdp and \HHOdpp fragmentation
and corresponding kinetic energy distributions (KEDs). The KEDs were obtained using 3D momentum
calibration. The green arrows show the ranges applied for the selection of the slow ions originating
from \HHOdp fragmentation.}
   \label{fig:rad}
\end{figure*}

\autoref{fig:rad} provides an overview of all analyzed ions resulting from \HHOdp and \HHOdpp
fragmentation. While the contribution from \HHOdp exhibits broad, "blob-like" features from 0~eV to
$\approx$1~eV, the contribution from the \HHOdpp Coulomb explosion displays narrow bands in the
\HHHOp, \HHOp, $\text{OH}^+$, and $\text{O}^+$ radial distributions with high kinetic energy centred
around $2$~eV due to the repulsion of charged fragments. The green arrows indicate the ranges used
to filter ions originating from the \HHOdp fragmentation.

In the analysis, we did not use the bottom part of some of the velocity map images due to a strong
background signal originating from the diffused water in the spectrometer chamber.

\subsection{Pump-probe signal of detected ion channels}
\label{sec:PP-signal}

\begin{figure*}
   \includegraphics[width=1.0\linewidth]{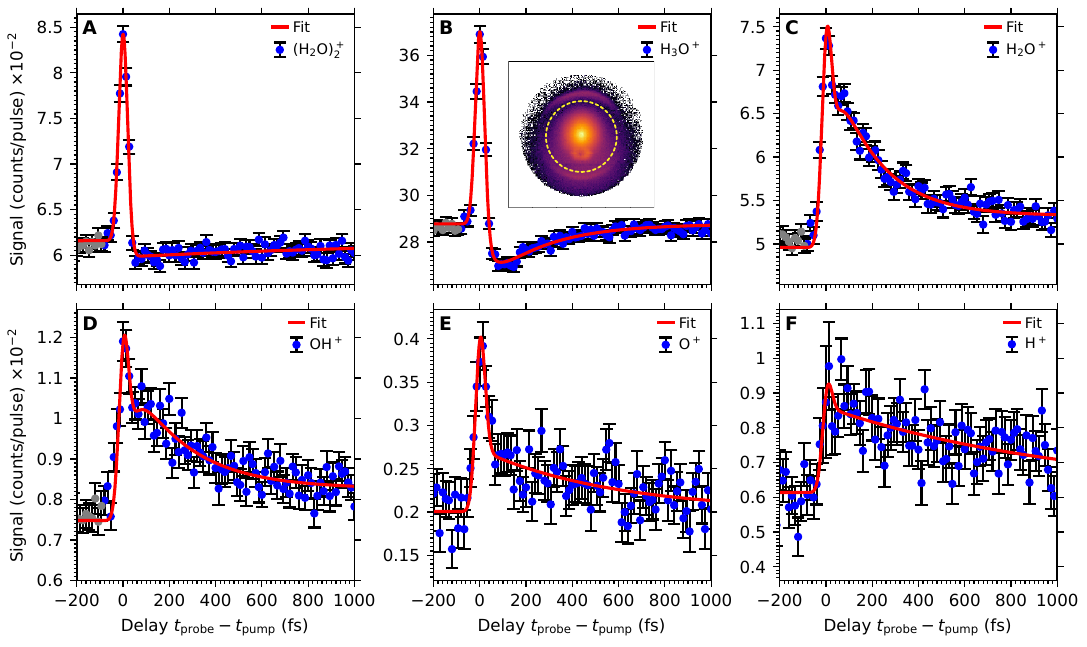}
   \caption{Transient signals of the most prominent ion channels. Panels A–F display \HHOdp, \HHHOp,
\HHOp, \OHp, \Op, and \Hp, respectively; see \autoref{sec:Overview} for details on detected species
and channel selection. Red traces represent fits using the parametric model based on
exponentially-modified-Gaussian (\emgf) functions described in~\autoref{sec:Methods}. The gray data
points in the range $(-500,-50)$~fs were excluded from the fitting procedure due to probe-induced
pre-excitation effects. The inset in panel B shows a velocity map of \HHHOp{}, whose central part,
originating from \HHOdp{} dynamics, is marked with a yellow-dashed circle.}
   \label{fig:PP}
\end{figure*}

Applying the regions of interest shown in \autoref{fig:rad}, we selected only ions originating from
\HHOdp fragmentation and plotted their pump--probe signal dependence in \autoref{fig:PP}. The signal
was corrected for molecular beam intensity as the data acquisition took several days. All channels
except for H$_3^+$ and H$_2^+$ show strong pump--probe dependence induced by disruptive probing
pulse. Notice that pump--probe signals of \HHOdp and \HHHOp show signal bleaching in t > 0~fs. This
reflects that the disrupting pulse is causing ion signal migration to other channels.

In this work, we focus on the signal of \HHOdp, \HHHOp, \HHOp, OH$^+$, O$^+$, and H$^+$. Transient
signals of these channels were also analysed by fitting the model based on a set of first-order
differential equations with a Gaussian function as a source described in~\autoref{eq:PP-model}.

The resulting fits are shown as red traces in~\autoref{fig:PP}, and the extracted parameters are
summarized in~\autoref{tab:PP-model-parameters}.
\begin{table*}[ht]
   \caption{Summary of all obtained parameters from the characterization of the pump--probe signals
presented in~\autoref{fig:PP} using the \emgf model described in \autoref{sec:Methods}.}%
      \label{tab:PP-model-parameters} \centering
      \begin{tabular}{c c c c c c c c c} \hline [Ch$^+$] & $a_{\text{IRF}}$ & $2.35 \cdot
\sigma_{\text{IRF}}$ & $a_1$ & $\tau_1$ & $a_2$ & $\tau_2$ & $a_\theta$ & [Ch$^+$]$_0$ \\ [0.5ex]
X$^+$ & [X$^+$]/pulse & fs & [X$^+$]/pulse & fs & [X$^+$]/pulse & fs & [X$^+$]/pulse &
[X$^+$]/pulse\\ \hline \HHOdp & 1.13(3) & 44.9(14) & -2.33(65) & 1307(445) & - & - & - & 0.0616(1)
\\ H$_3$O$^+$ & 4.28(11) & 46.8(11) & -6.97(39) & 250(28) & 1.09(38) & 39(9) & - & 0.2877(2) \\
H$_2$O$^+$ & 1.01(6) & 48.6(0) & 3.88(22) & 199(24) & -0.51(17) & 26(6) & 0.00366(32) & 0.0496(2) \\
OH$^+$ & 0.19(2) & 48.6(44) & 0.78(12) & 249(75) & -0.10(7) & 33(15) & 0.00076(17) & 0.0075(1) \\
O$^+$ & 0.09(1) & 48.6(16) & 0.41(6) & 578(119) & - & - & - & 0.0020(0) \\ H$^+$ & 0.08(2) &
47.0(94) & 2.60(38) & 1056(194) & - & - & - & 0.0061(1) \\ [1ex] \hline
      \end{tabular}
\end{table*} While the \HHOdp, \Hp, and \Op traces are well described by a single-term \emgf model,
the other ion species presented in~\autoref{fig:PP} require a double-term \emgf model.

\subsection{DP signal correlations}
\label{sec:DP-correlations}

\begin{figure}
   \includegraphics[width=0.85\linewidth]{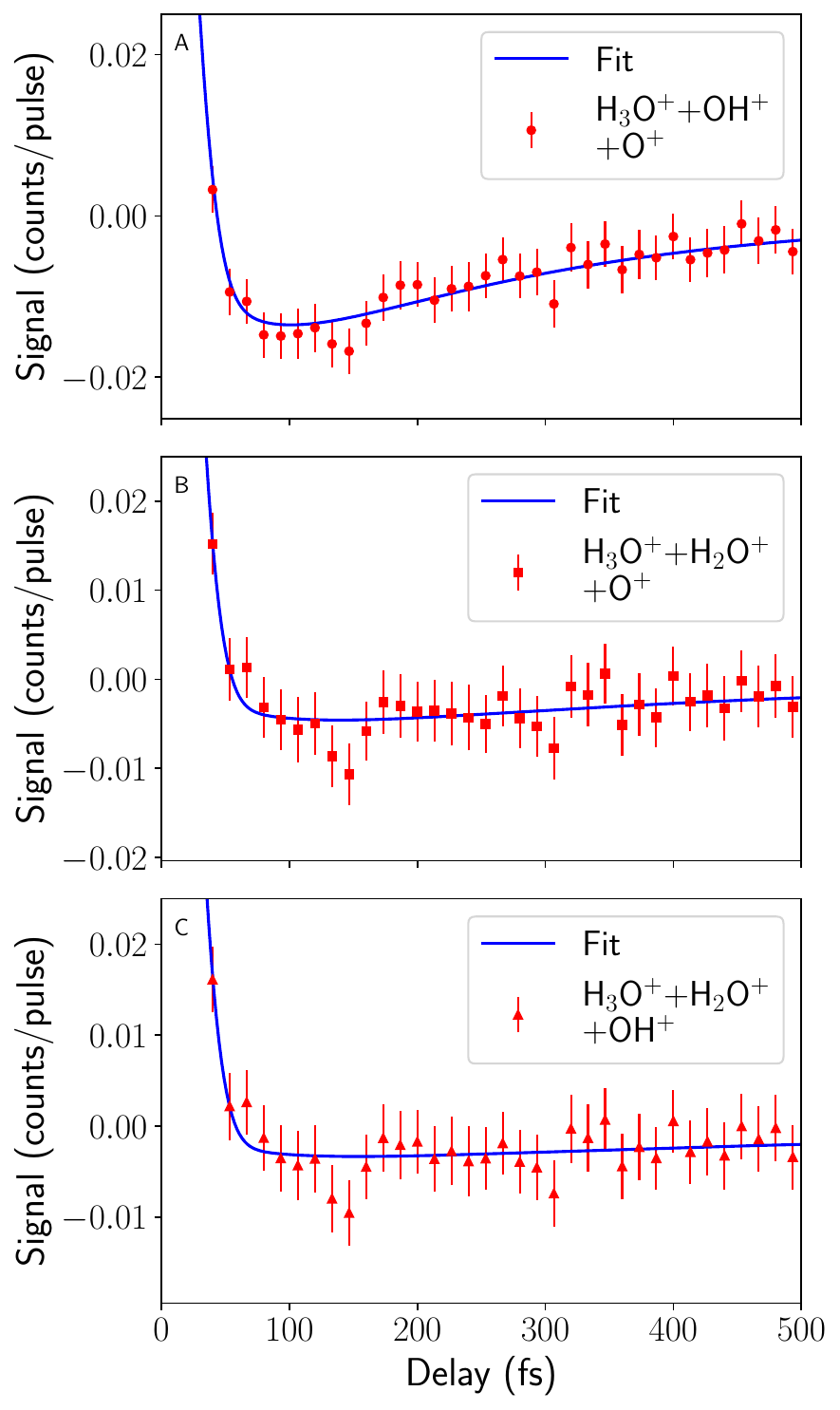}%
   \caption{Summed transient signals of various ion channels with the fitted parametric model
described by \autoref{eq:PP-model}.}
   \label{fig:pp-summed}
\end{figure}

The disruptive probing field perturbs the ongoing fragmentation dynamics in \HHOdp, resulting in a
redistribution of the ion signal across different ion channels. This redistribution can be tracked
by analyzing correlations among the transient signals of individual ion species. In the most
straightforward approach, such correlations can be examined without utilizing the kinetic energy
information.

A clear correlation is observed between the transient signals of \HHOdp and \Hp, whose summation
yields a flat pump--probe trace. This behavior is also reflected in the fitted amplitudes, which
exhibit equal magnitudes but opposite signs, and in the characteristic time constants on the order
of $\sim1$~ps. In a similar fashion, we investigated the correlation between the bleaching of the
\HHHOp transient signal and the rising signals of the \HHOp, \OHp, and \Op channels. To this end,
pairs of \HHOp, \OHp, and \Op transients were summed with the \HHHOp signal, and the resulting
traces were fitted using the \emgf model described in \autoref{eq:PP-model}.

\begin{table}[htbp] \centering
\caption{Comparison of the fitted lifetimes of summed transient signals of various ion channels and
the correlated ion channels.}
\begin{tabular}{|c|c||c|c|} \hline Summed channels & $\tau_1$ (fs) & Channel & $\tau_1$ (fs) \\
\hline \HHHOp + \HHOp + \OHp & 550(264) & \Op & 578(119) \\ \HHHOp + \HHOp + \Op & 365(131) & \OHp &
386(101) \\ \HHHOp + \OHp + \Op & 229(32) & \HHOp & 199(24) \\ \hline
\end{tabular}
\label{tab:pp-summed-correlation}
\end{table}

The resulting transient signals and their corresponding fits are shown in~\autoref{fig:pp-summed}.
The extracted characteristic times are in excellent agreement with the time constants of the omitted
ion channel, confirming the underlying correlations. All fitted time constants are summarized
in~\autoref{tab:pp-summed-correlation}.

\begin{figure}
      \includegraphics[width=0.8\linewidth]{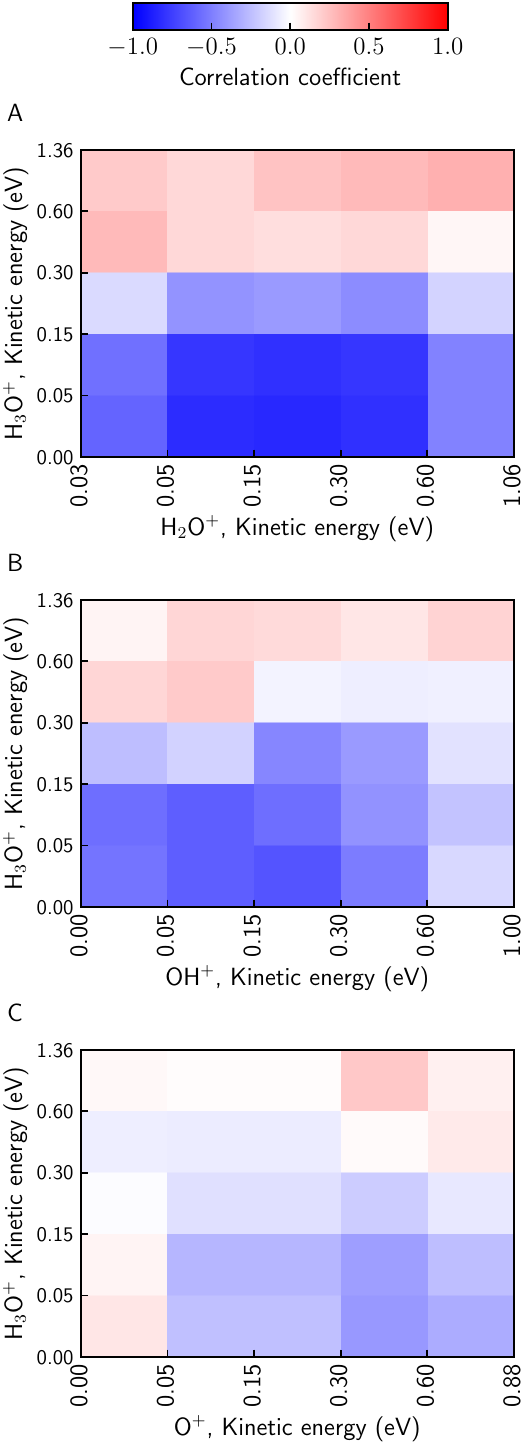}%
      \caption{Energy-resolved correlation matrices of \HHHOp with (A) \HHOp, (B) \OHp, and (C)
\Op.}
      \label{fig:PP-correlations}
\end{figure}

So far, the kinetics were analyzed for individual ion species across all kinetic energies, and the
kinetic energy information obtained from the VMI approach was primarily used to filter out ions
originating from Coulomb explosion channels. In this section, we extend the use of kinetic energy
information assigned to each detected ion to further decompose the acquired transient signals.

The kinetic energy range for each ion species resulting from \HHOdp fragmentation was divided using
non-linear binning. We applied the \emgf model to fit the transient signals and remove
contributions from the cross-correlation and constant background terms. After background
subtraction, we performed Pearson correlation analysis on the cleaned transient signals. The results
of this analysis are presented in~\autoref{fig:PP-correlations} for \HHHOp with other species.

\autoref{fig:PP-correlations} reveals the redistribution of ion signal from the \HHHOp channel to
the \HHOp, \OHp, and \Op channels. Strong anti-correlation (blue regions) indicates signal depletion
in \HHHOp accompanied by signal increase in the correlated ion channels, representing transitions
due to disruptive probing. Conversely, positive correlation (red regions) suggests that multiple ion
species may originate from a shared parent pathway and exhibit similar dynamic behavior.

Specifically, \HHHOp ions with kinetic energy below 0.15~eV correlates negatively with the
population of \HHOp ions in the (0.05, 0.6)~eV range, and shows weaker anti-correlation with \OHp
ions up to 0.3~eV and \Op ions between (0.3, 0.6)~eV. For \HHHOp ions in the (0.15, 0.3)~eV range,
similar levels of anti-correlation appear with \HHOp (0.05, 0.6)~eV and \OHp (0.15, 0.6)~eV, along
with a minor contribution from \Op (0.3, 0.6)~eV. Finally, \HHHOp ions with kinetic energies between
(0.3, 0.6)~eV that are bleached by the disruptive pulse are redistributed into \OHp ions in the
(0.15, 1)~eV range and \Op ions in the (0, 0.3)~eV range. These energy-resolved correlations of
various species with \HHHOp transient signal highlight the various impacts of the disruptive field
on the ongoing fragmentation dynamics.

\subsection{Pump-probe dependence of ions' kinetic energy distributions} 
\begin{figure*}
   \includegraphics[width=0.70\linewidth]{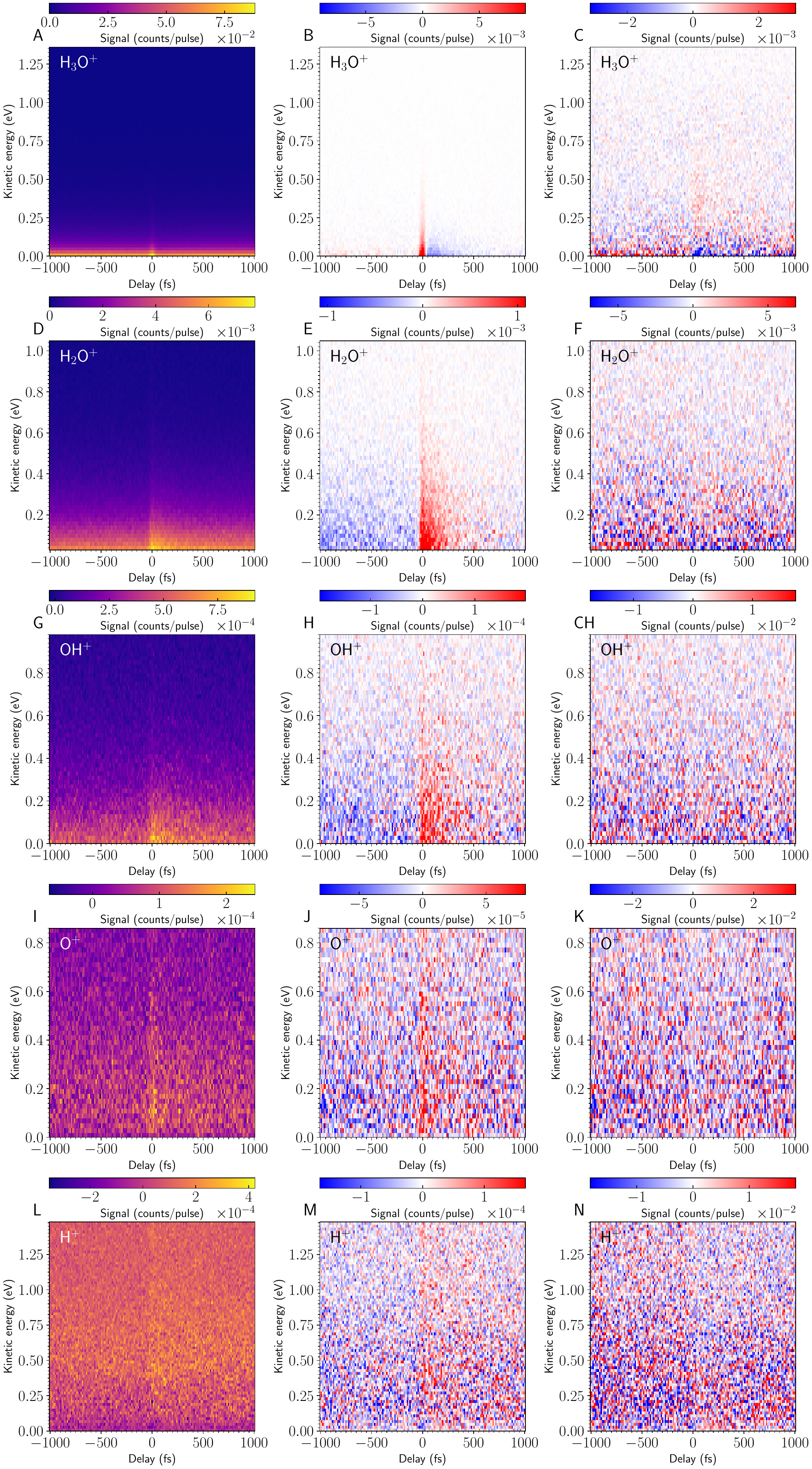}%
   \caption{Ion maps representing the energy-resolved pump--probe dependence of the individual ion
channels. The first column corresponds to the absolute signal, the second column shows relative
changes from pump-only signal level, and the last column shows relative changes normalized in each
row.}
   \label{fig:PP-maps}
\end{figure*}

The ion maps shown in
\autoref{fig:PP-maps} illustrate the pump--probe dependence of the ion signal as a function of
kinetic energy. The first column corresponds to the absolute signal, the second column shows
relative changes, and the last column shows relative changes normalized in each row.

\subsection{Kinetic-energy-resolved signal}
\label{sec:hydroinum-SI}

\begin{figure*}
      \includegraphics[width=1.0\linewidth]{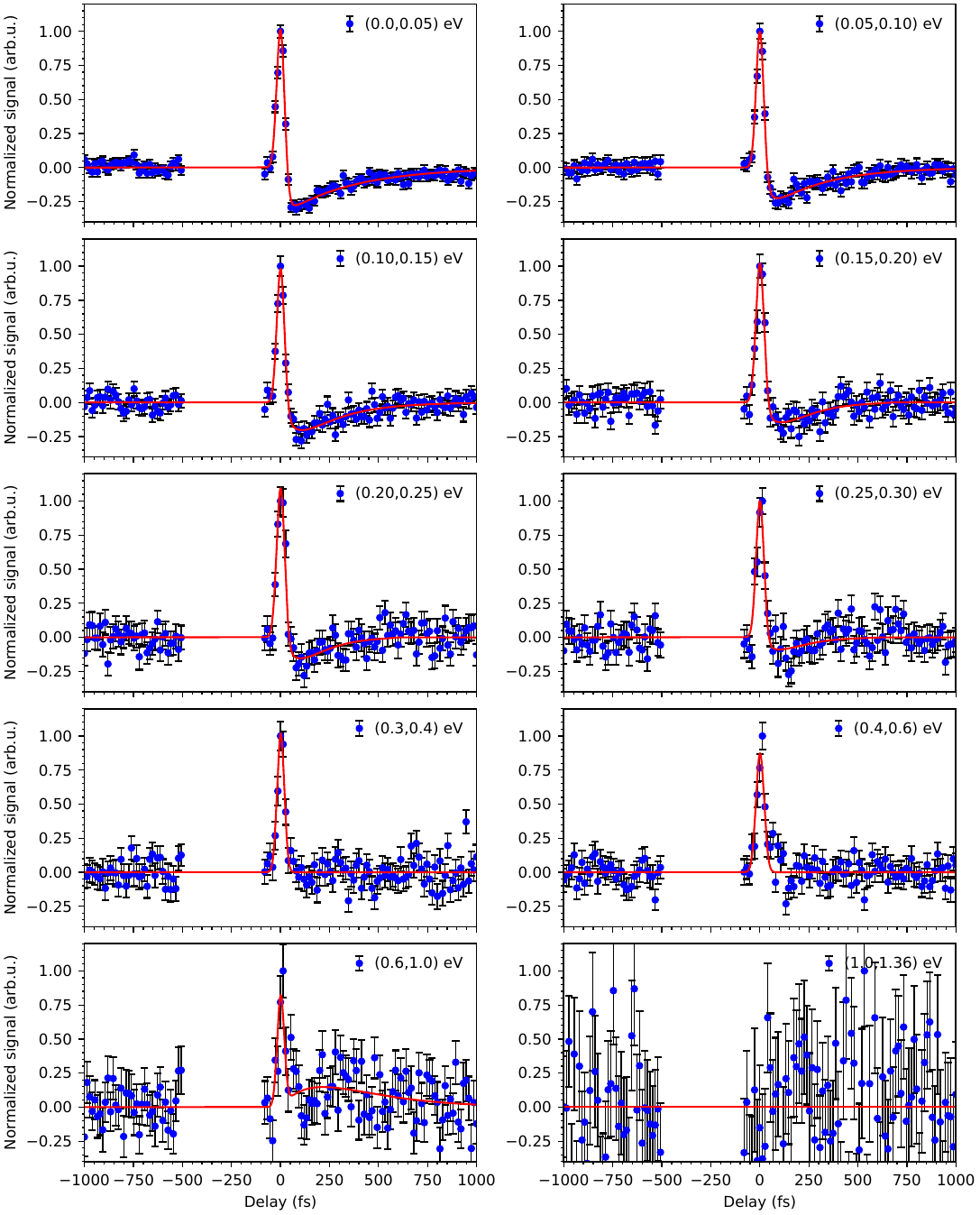}%
      \caption{Transient signals of \HHHOp{} at selected kinetic energy ranges. The data were fitted
using the parametric \emgf model described in \autoref{sec:Methods}. }
      \label{fig:Lifetimes-H3O}
\end{figure*}

Figure~2 in the main text shows results of analysis of
kinetic-energy-resolved \HHHOp transient signal. A two-dimensional ion-signal histogram of \HHHOp
with a sudden signal enhancement at time zero in red, corresponding to the cross-correlation peak,
and the signal depletion at positive delays in blue is displayed in
\autoref[B]{fig:PP-maps}. Analogous to the individual-ion-channel analysis, we applied the parametric \emgf
model across ten predefined kinetic-energy ranges up to 1.36~eV; this upper limit was chosen to
avoid overlap with the Coulomb-explosion channel, as illustrated in \autoref{fig:rad} and in the inset of
\autoref[A]{fig:PP}.
We observed distinct time-dependences among the individual datasets corresponding to
the specific kinetic-energy release. All fits are shown in \autoref{fig:Lifetimes-H3O}.
For instance, the top dataset obtained from \HHHOp{} ions with
KE in the range of (0.0,0.05)~eV shows signal bleach following the cross-correlation signal in
contrast with the bottom dataset obtained for \HHHOp{} ions with KE in the range of (0.4, 0.6)~eV.
See also the main text for a detailed discussion about the energy-resolved observations made from \HHHOp{} data.

\begin{figure}
      \includegraphics[width=0.8\linewidth]{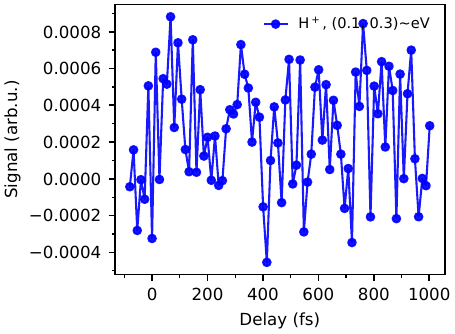}%
      \caption{Oscillatory component of the \Hp ion signal in the kinetic energy range of 0.1--0.3~eV.}
      \label{fig:H-oscillations}
\end{figure}

Similar to the \HHHOp{} case, we obtain an energy-resolved map of the \Hp{} transient signal
after subtracting pump-only contributions shown in Figure~3 in the main text. Most of the redistributed \HHOdp{} signal appears in \Hp{} ions with kinetic energies
between 0.3 and 1.5~eV, with lifetimes of $0.7$–$1.4$~ps. In contrast, we observe a rapid bleach
within the first 100~fs at kinetic energies below 0.1~eV, which cannot be attributed to a depletion
of cross-correlation. Notably, in the 0.1–0.3~eV range, the transient \Hp{} signal exhibits an
oscillatory modulation with a period of approximately 250~fs, as illustrated
in~\autoref{fig:H-oscillations}. The origin of the oscillations remains unresolved.

\subsection{\emph{Ab initio} electronic-state resolved trajectory simulations}
\label{sec:Simulations}

The interpretation of the experimental data is facilitated by \textit{ab initio} calculations. As in
earlier studies~\cite{Loh:Science367:179,Li:Science383:1118}, we populated \HHOdp{} in experiment
via strong-field ionization, raising the question of which electronic states drive its dynamics. Our
recent study~\cite{Vinklarek:water2-moadk:inprep} showed that ionization predominantly accesses the
four energetically lowest states, \Dzero{}–\Dthree{}, which \emph{ab initio}
simulations~\cite{Schnorr:SciAdv9:7864} found to undergo proton transfer on distinct sub-100~fs
timescales. Across all four considered electronic states, the primary dynamical process is ultrafast
proton transfer, forming the ion–radical complex $[\HHHO{}\cdots\text{OH}]^+$. This complex either
stabilizes, a channel observed dominantly in the \Dzero{} state, or fragments to yield a separated
\HHHOp{} ion and OH radical. Here, we extend the simulations to 1~ps, allowing us to resolve
fragmentation pathways beyond the initial PT event.

\begin{figure}
   \includegraphics[width=1.0\linewidth]{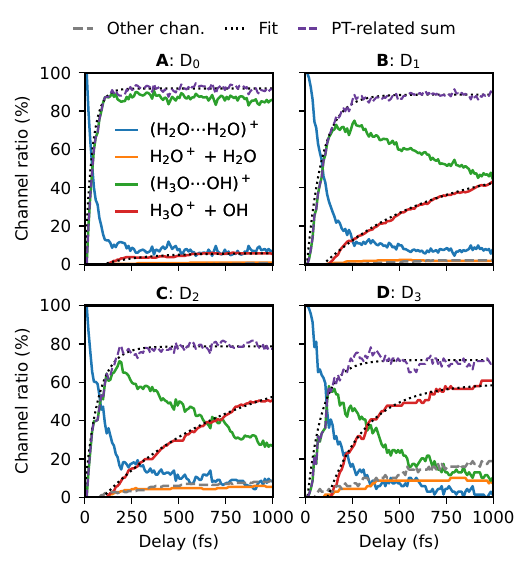}%
   \caption{\emph{Ab initio} trajectory simulations of \HHOd{} dynamics after single ionization into
the four lowest electronic states (D$_0$–D$_3$). The curves show the time-dependent relative
populations of the main ion–radical channels: $[\HHO\cdots\HHO]^+$ (blue), \HHOp + H$_2$O
(orange), $[\HHHO{}\cdots\text{OH}]^+$ (green), and \HHHOp + OH (red), with the cumulative contribution
of the remaining minor channels shown as a gray dashed line. Channels attributed to proton transfer
(dashed violet) and fragmentation into \HHHOp + OH were fitted with single-exponential models (black
dotted).}
   \label{fig:Trajectory-simulations}
\end{figure}

\autoref{fig:Trajectory-simulations} provides simulated time-dependent relative branching into the
most dominant channels initialized at the four electronic states, \Dzero{}--\Dthree.
 Our simulations show that a portion of $[\HHHO{}\cdots\text{OH}]^+$ complexes remain intact
to 1~ps. A smaller fraction of trajectories, indicated by blue lines, on the \Dzero{}--\Dtwo{}
states yield a weakly bound $[\HHO\cdots\HHO]^+$ complex with hemi-bonded
structure~\cite{Chipman:JPCA120:9618,Iguchi:JPCL14:8199,Hartweg:JPCA125:4882}. Here, the values
converged to non-zero channel ratios at 1~ps. In addition, the \Dtwo{} and \Dthree{} states display
enhanced fragmentation into \HHOp{} + \HHO, indicated by orange lines, compared to the other states.
Furthermore, less common fragmentation pathways -- including \OHp{} formation, proton ejection, and
stabilization of more complex ionic species such as H$_2$O$_2^+$ -- become increasingly prominent
with increasing ionization energy, rising from below 2\% at \Dzero{} to about 20\% in \Dthree{} at
the time of 1~ps. Single-exponential fits to the $[\HHHO{}\cdots\text{OH}]^+$ formation and \HHHOp
fragmentation channels yield lifetimes of the proton migration $\tau_{\text{PT}}$ and fragmentation
$\tau_{\HHHOp}$, see \autoref{tab:Lifetimes-simulated}, which characterize the averaged early-phase
post-ionization dynamics of individual electronic states.
\begin{table} \centering
   \caption{Characteristic $1/e$-timescales $\tau_\text{PT}$ and $\tau_{\HHHOp}$ from trajectory
simulations, describing $[\HHHO{}\cdots\text{OH}]^+$ formation via proton transfer and subsequent
fragmentation into \HHHOp across different electronic states. The applied parametric model is
discussed in \autoref{sec:Simulation-par-model}.}
   \label{tab:Lifetimes-simulated}
   \begin{tabular}{ccccc} \hline State & \Dzero & \Done & \Dtwo & \Dthree \\ \hline
$\tau_{\text{PT}}$~(fs) & 38(1) & 85(2) & 71(2) & 81(4) \\ $\tau_{\HHHOp}$~(fs) & 259(19) & 413(10)
& 725(31) & 212(6) \\ \hline
   \end{tabular}
\end{table}

\subsection{Fitting of pump--probe ion signal}
\label{sec:PP-Fitting}

The applied model described by~\autoref{eq:PP-model} is result of the convolution of IRF function
with a delta function $\delta$ representing the initial source and cross-correlation signal,
summation over various exponential functions reflecting the investigated dynamics, and heavy-side
function corresponding to the secondary fragmentation of previously stabilized ion fragments
\begin{equation}
      \label{eq:Model-SI-convolution}
      \begin{aligned} \left[\text{Ch}^+\right] & \left(t; a_{\text{IRF}}, \sigma_{\text{IRF}},
\{a_k\}, \{\tau_k\}, a_{\theta} \right) = \\ g & \left(t; a_{\text{IRF}}, \sigma_{\text{IRF}}
\right) * \\ & \left(\delta(t) + \sum_{k} \exp_k\left(t; a_k, \tau_k\right) + \Theta\left(t;
a_{\theta} \right) \right),
      \end{aligned}
\end{equation} where we omitted the constant term [Ch$^+$]$_0$, which is not connected to the
searched dynamics. \autoref{eq:Model-SI-convolution} reflects that our model in the main text is
based on first-order linear differential equations with a Gaussian function as a source.

\begin{table*} 
   \caption{Summary of all obtained parameters from the fitting of the pump--probe signal when using
the model described by~Equation~1 in the main text with varying $t_0$ and $\sigma_{\text{IRF}}$
parameters. }
   \label{tab:pp-fits-SI} \centering
   \begin{tabular}{c c c c c c c c c c} \hline ion & $a_{\text{IRF}}$ & $t_0$ & $2.35 \cdot
\sigma_{\text{IRF}}$ & $a_1$ & $\tau_1$ & $a_2$ & $\tau_2$ & $a_\theta$ & $I_0$ \\ [0.5ex] X$^+$ &
[X$^+$]/pulse & fs & fs & [X$^+$]/pulse & fs & [X$^+$]/pulse & fs & [X$^+$]/pulse & [X$^+$]/pulse \\
\hline \HHOdp & 1.12(3) & 2.37(61) & 44.4(14) & -2.26(55) & 1181(356) & - & - & - & 0.0616(1) \\
H$_3$O$^+$ & 4.34(13) & 3.02(57) & 47.0(11) & -6.88(36) & 268(26) & 0.69(28) & 27(9) & - & 0.2879(2)
\\ H$_2$O$^+$ & 0.86(12) & -5.81(243) & 48.8(39) & 3.95(21) & 205(20) & -0.36(16) & 18(7) &
0.00376(29) & 0.0495(1) \\ OH$^+$ & 0.20(3) & -3.45(404) & 64.0(75) & 0.78(11) & 245(77) & -0.11(9)
& 35(21) & 0.00076(17) & 0.0074(1) \\ O$^+$ & 0.10(1) & 3.17(312) & 62.4(70) & 0.42(7) & 632(144) &
- & - & - & 0.0020(0) \\ H$^+$ & 0.10(6) & -2.10(933) & 43.5(162) & 2.54(36) & 986(191) & -0.04(5) &
15(20) & - & 0.0061(1) \\ \hline
   \end{tabular}
\end{table*}

In the main text, we apply an additional constraint on the amplitudes of \emgf and rate constants for
observed dynamics with double \emgf shape
\begin{equation}
\label{eq:condition-SI} a_1\cdot k_1 = - a_2 \cdot k_2.
\end{equation} We derived this condition based on the assumption that the final ion product is
populated from a metastable state. The differential equations describing such a scenario are
\begin{align} \odv{[1]}{t} = g(t) - k_{1\rightarrow2} [1] - k_{1\rightarrow4} [1], \label{eq:diff-1}
\\ \odv{[2]}{t} = + k_{1\rightarrow2}[1] - k_{2\rightarrow3}[2], \label{eq:diff-2} \\ \odv{[3]}{t} =
+ k_{2\rightarrow3}[2], \label{eq:diff-3}
\end{align} where we marked the population of states as $[x],~x \in \{1,2,3\}$ and $k_{x\rightarrow
y}$ are the rate constants, \ie, $k_{x\rightarrow y} >0$. Furthermore, assuming zero population of
$[1]$ in negative times, $t<0$, we obtain a solution for [2] state population equal to
\begin{align}
\label{eq:solution-2-SI}
\begin{split} [2] (t; k_{1\rightarrow2},k_{2\rightarrow3}, k_{1\rightarrow4}, \sigma) = & \\
\frac{k_{1\rightarrow2}}{2(k_{1\rightarrow2}+ k_{1\rightarrow4} - k_{2\rightarrow3})} & \\ \times [
g(t; k_{2\rightarrow3}, \sigma) - & g(t; k_{1\rightarrow2} + k_{1\rightarrow4}, \sigma) ],
\end{split}
\end{align} where the decay function \( g_k(t; \sigma) \) is defined as
\begin{equation}
\label{eq:g(t)-SI} g(t; k, \sigma) = e^{-k t + \frac{1}{2} k^2 \sigma^2} \left(1 +
\operatorname{Erf}\left( \frac{t - k \sigma^2}{\sqrt{2} \sigma} \right) \right).
\end{equation} The formula for \emgf is equal to
\begin{equation}
\label{eq:emg-general-SI} \emgf(t; a, k, \sigma) = \frac{a\cdot k}{2}g(t; k, \sigma),
\end{equation} where $a$ is amplitude of \emgf. Through comparison of \autoref{eq:solution-2-SI},
\autoref{eq:g(t)-SI}, and \autoref{eq:emg-general-SI}, we obtain the additional constraint
\autoref{eq:condition-SI} for fitting dynamics with the double \emgf model.

For a comparison of the fitting results presented in Table~1 in the main text using non-varying
$t_0$ and upper limit $\sigma^{\text{max}}_{\text{IRF}}$, we also show the estimated fitting
parameters in case that both the parameters were set free in~\autoref{tab:pp-fits-SI}. This
illustrates the broad range of obtained values of $t_0$ and $\sigma_{\text{IRF}}$, see \eg, \HHOp,
\HHHOp, and OH$^+$ channels for comparison. We assume that the broad widths characterized by
$\sigma_{\text{IRF}}$ in the range from 60~fs originate from the underlying fast dynamics.
Therefore, in the fitting presented in the main text we applied an upper limit for
$\sigma_{\text{IRF}}$ to reveal the fast dynamics occurring on the tenths of femtosecond timescale.

Our model is based on a kinetic model similar to \cite{Jochim:RSI93:033003}. To strengthen the
validity of our fitting model and its agreement with the experimental data, we also checked the
p-values for all fits. Conducting a Kolmogorov-Smirnov test on the obtained distribution of p-values
reveals that it is consistent with the statistically expected uniform distribution over the interval
$[0,1]$. The obtained p-value for the Kolmogorov-Smirnov test is 0.16, falling near the center of
the expected D-distribution for $n=22$ -- the critical value for $\alpha=0.05$ here is 0.28. This
indicates that our empirical model does not contradict the data.

\onecolumngrid
\clearpage
\twocolumngrid

\section{Tutorial on disruptive probing}
\label{sec:tutorial}
\subsection{Motivation}
\label{sec:disprob-motivation}
A recently published paper by Jochim et al.~\cite{Jochim:RSI93:033003} presents a new holistic
pump--probe approach called disruptive probing, which can be highly beneficial for the simultaneous
tracking of multiple fragmentation pathways after target ionization.

Unfortunately, the description of the time-dependent probing-pulse-induced dynamics, whose effect
should reflect the searched original dynamics, was described rather vaguely on the side of the
presented description by the kinetics model represented by a set of differential equations.

The following chapters describe seven example situations that illustrate the process of disruptive
probing and offer clear insight into the origins of the observed signal dependencies using the
differential equation ansatz. This is particularly beneficial for the analysis and subsequent
interpretation of the measured data discussed in the main article.

\subsection{Example 1: 3 level model}
\label{sec:disprob-example1}

A model compound $\text{ABC}$, consisting of three distinguishable parts $\text{A}$, $\text{B}$, and
$\text{C}$, is ionized into an initial state $\text{ABC}^{+*}$, denoted as $\text{i}_0$. The
compounds of this state can either fragment into $\text{AB}^{+}+\text{C}$ with a rate constant
$k_{\text{i}_0\rightarrow\text{AB}^{+}}$ or stabilize as the $\text{ABC}^{+}$ ion with a rate
constant $k_{\text{i}_0\rightarrow\text{ABC}^{+}}$. These processes can be expressed through the
following differential equations:
\begin{align} \odv{[\ip]}{t} =& g(t) - k_{\text{i}_0\rightarrow\ABp} [\ip] -
k_{\text{i}_0\rightarrow\ABCp} [\ip], \label{eq:dif1-ip}\\ \odv{[\ABCp]}{t} =& +
k_{\text{i}_0\rightarrow\ABCp} [\ip], \label{eq:dif1-ABC} \\ \odv{[\ABp]}{t} =& +
k_{\text{i}_0\rightarrow\ABp} [\ip], \label{eq:dif1-AB}
\end{align} where $g(t)$ is an arbitrary function describing the \ip population by the ionizing
laser pulse. In the paper~\cite{Jochim:RSI93:033003}, they apply a Gaussian function $g(t) =
\exp\left(-\frac{t^2}{2\sigma_t^2}\right)$ as an approximation of the laser pulse temporal profile,
where $\sigma_t = \text{FWHM}_t/2.355$ is the standard deviation. In the simulations below, we also
apply a Gaussian function, including the proper normalization factor to ensure the correct units of
$\text{s}^{-1}$ in
\begin{equation} g(t) = \frac{1}{\sqrt{2\pi \sigma_t^2}}\exp{\left(-\frac{t^2}{2 \sigma_t^2}
\right)}
\end{equation}

\begin{figure}
   \includegraphics[width=\linewidth]{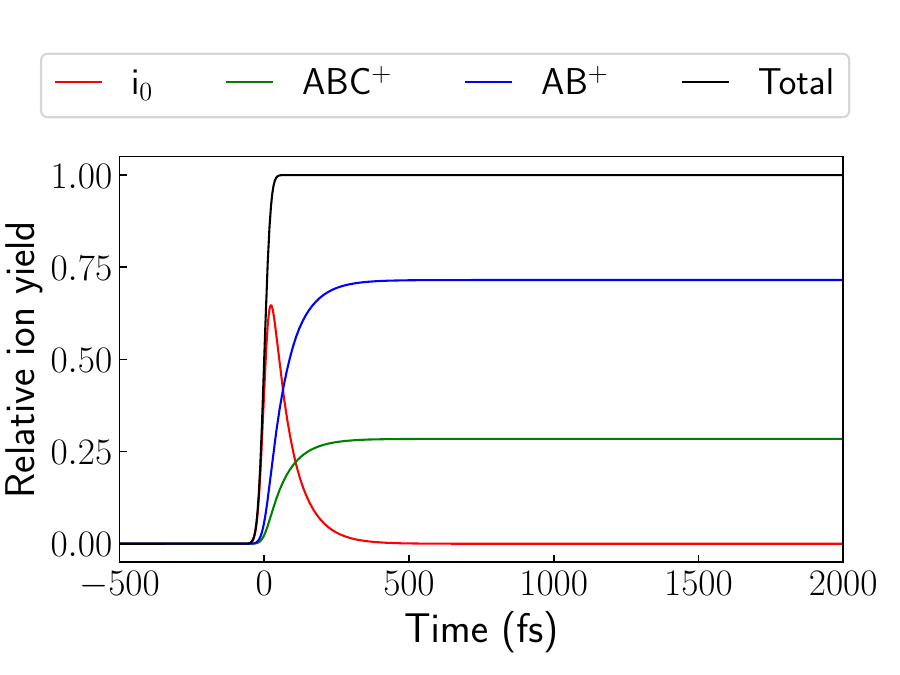}%
   \caption{Dynamics of the ABC compound after ionization according to Example~1.}
   \label{fig:Dynamics-1}
\end{figure}

The analytic solution of the first differential equation \autoref{eq:dif1-ip} leads to an
exponentially-modified-Gaussian function
\begin{align}
   \begin{split} [\ip](t,A, \mu, \sigma_t, k_{\ip}) = & \frac{A k_{\ip}}{2} \\ & \times
\exp{k_{\ip}\left(\frac{k_{\ip} \sigma_t^2}{2}-t \right)} \\ & \times \text{erfc}
\left(\frac{k_{\ip}\sigma_t^2 - t}{\sqrt{2}\sigma_t}\right), \label{eq:exp-gauss}
   \end{split}
\end{align} where $A$ is the amplitude and $k_{\ip}$ is the summed rate constant for \ip decaying
into other levels. In the case above, $A = 1$ and $k_{\ip} = k_{\ip\rightarrow\ABCp} +
k_{\ip\rightarrow\ABp}$.

Alternatively, by numerically solving the set of differential
equations~\ref{eq:dif1-ip}–\ref{eq:dif1-AB} with the parameters $\sigma_t = (40~\text{fs})/2.355 =
17~\text{fs}$, $k_{\ip \rightarrow \ABp} = 1/(100~\text{fs})$, and $k_{\ip \rightarrow \ABCp} =
1/(250~\text{fs})$, we obtain the signal dependencies for \ip, \ABp, and \ABCp shown in
\autoref{fig:Dynamics-1}. The goal of our experiment is to measure these dependencies. Since
fragmentation occurs on the subpicosecond timescale, it has practically no effect on the observed
time-of-flight mass spectra and velocity map images (mass spectrometric features assigned to
metastable ions would exhibit particular features from the nanosecond timescale of fragmentation),
we want to apply a disruptive probing technique to obtain them.

\subsubsection{Simulation of disruptive probing}
\label{sec:disprob-simulation}

\begin{figure}
   \includegraphics[width=\linewidth]{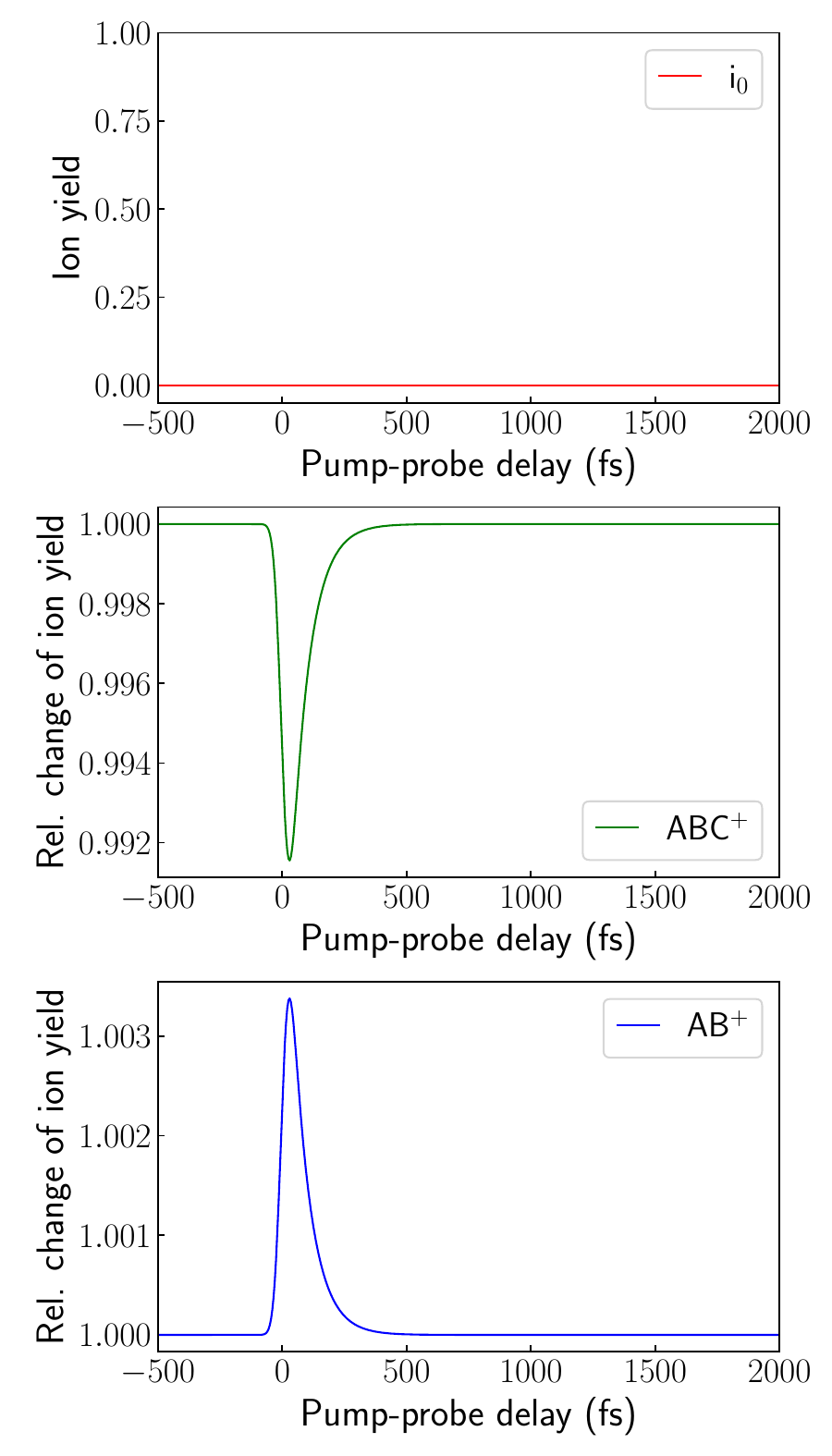}%
   \caption{Simulated disruptive probing measurement in case of Example~1.}
   \label{fig:Model-1}
\end{figure}

In disruptive probing, a second weak probing pulse interacts with an ensemble of ionized particles
undergoing relaxation processes. The probing pulse should be weak enough to affect the ongoing
processes, such as changes in couplings among the states and rate constants through changes in
potential energy surfaces, but not change the ion total population. In our simplified model, the
effect of the probing pulse is described by adding new or altered rate constants that represent the
weak-probe-pulse-induced couplings. We assume, as in the undisturbed model above, that the rate
parameters are constants. To simplify the notation, we incorporate the probe effect on couplings
directly into the multiplicative factor of the rate parameters as $(1 + \delta_{X^+ \rightarrow Y^+}
h(t))$, where $\delta_{X^+ \rightarrow Y^+}$ represents the relative change induced by the probe
pulse field and $h(t)$ is a function describing the probing pulse centered at the pump--probe delay
$\tau$, i.e., $h(t) = \frac{1}{\sqrt{2\pi \sigma_t^2}} \exp\left(-\frac{(t-\tau)^2}{2
\sigma_t^2}\right)$. Here, $\sigma_t$ of the probing pulse is taken to be equal to the standard
deviation of the ionizing pulse. This is a reasonable constraint, as in our experiment, the probing
pulse is generated by a beam splitter reflecting a fraction of the pumping pulse. Applying the
changes due to the probing pulse, we rewrite Equations~\ref{eq:dif1-ip}–\ref{eq:dif1-AB} as follows:
\begin{align}
   \begin{split} \odv{[\ip]}{t} =& g(t) \\ & - k_{\text{i}_0\rightarrow\ABp} (1 +
\delta_{\ip\rightarrow \ABp}h\left(t\right))[\ip] \\ & - k_{\text{i}_0\rightarrow\ABCp} (1 +
\delta_{\ip\rightarrow \ABCp}h\left(t\right))[\ip] \label{eq:m1-ip}
   \end{split} \\ \odv{[\ABCp]}{t} =& + k_{\text{i}_0\rightarrow\ABCp}(1 + \delta_{\ip\rightarrow
\ABCp}h\left(t\right))[\ip], \label{eq:m1-ABC} \\ \odv{[\ABp]}{t} =& +
k_{\text{i}_0\rightarrow\ABp}(1 + \delta_{\ip\rightarrow
\ABCp}h\left(t\right))[\ip]. \label{eq:m1-AB}
\end{align} The analytic solution of the new set of differential equations is not straightforward,
so we will use a numerical solver to simulate the results of the disruptive probing measurement. In
the real experiment, we measure the resulting ion yields of different ion fragments at different
pump--probe delays. Similarly, by varying the probe delay in the simulation, we can compare the
resulting ion populations at a later time, specifically at 3~ps when all fragmentation dynamics are
finished, to the ion yields in undisturbed fragmentation (static measurement without the probing
pulse). This is shown in \autoref{fig:Model-1}.

\begin{figure}
   \includegraphics[width=\linewidth]{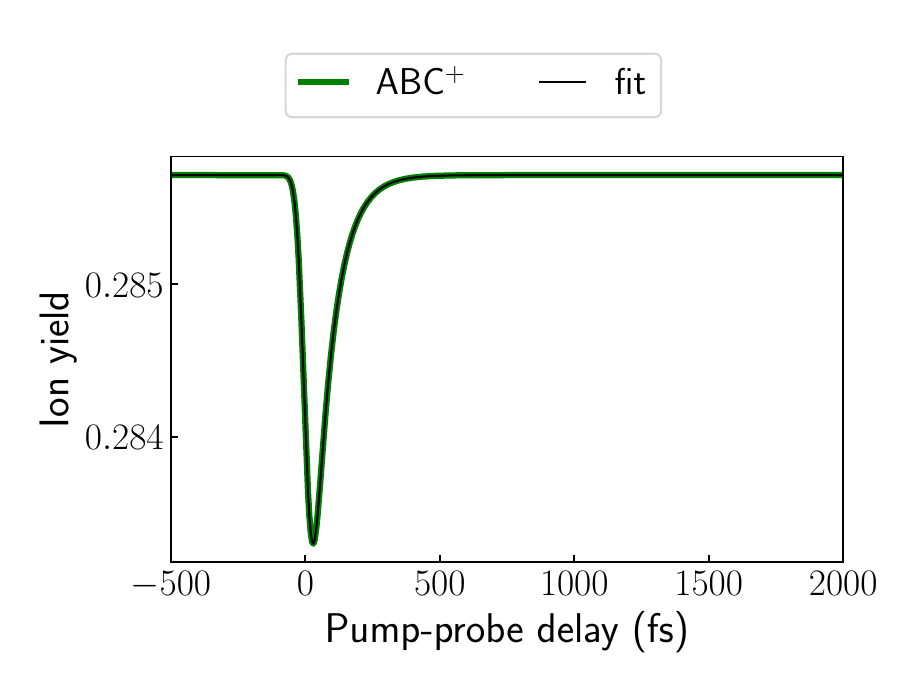}%
   \caption{Fitted simulated signal dependence of ABC$^+$ ion from Example~1.}
   \label{fig:Fitting-1}
\end{figure}

\begin{figure}
   \includegraphics[width=\linewidth]{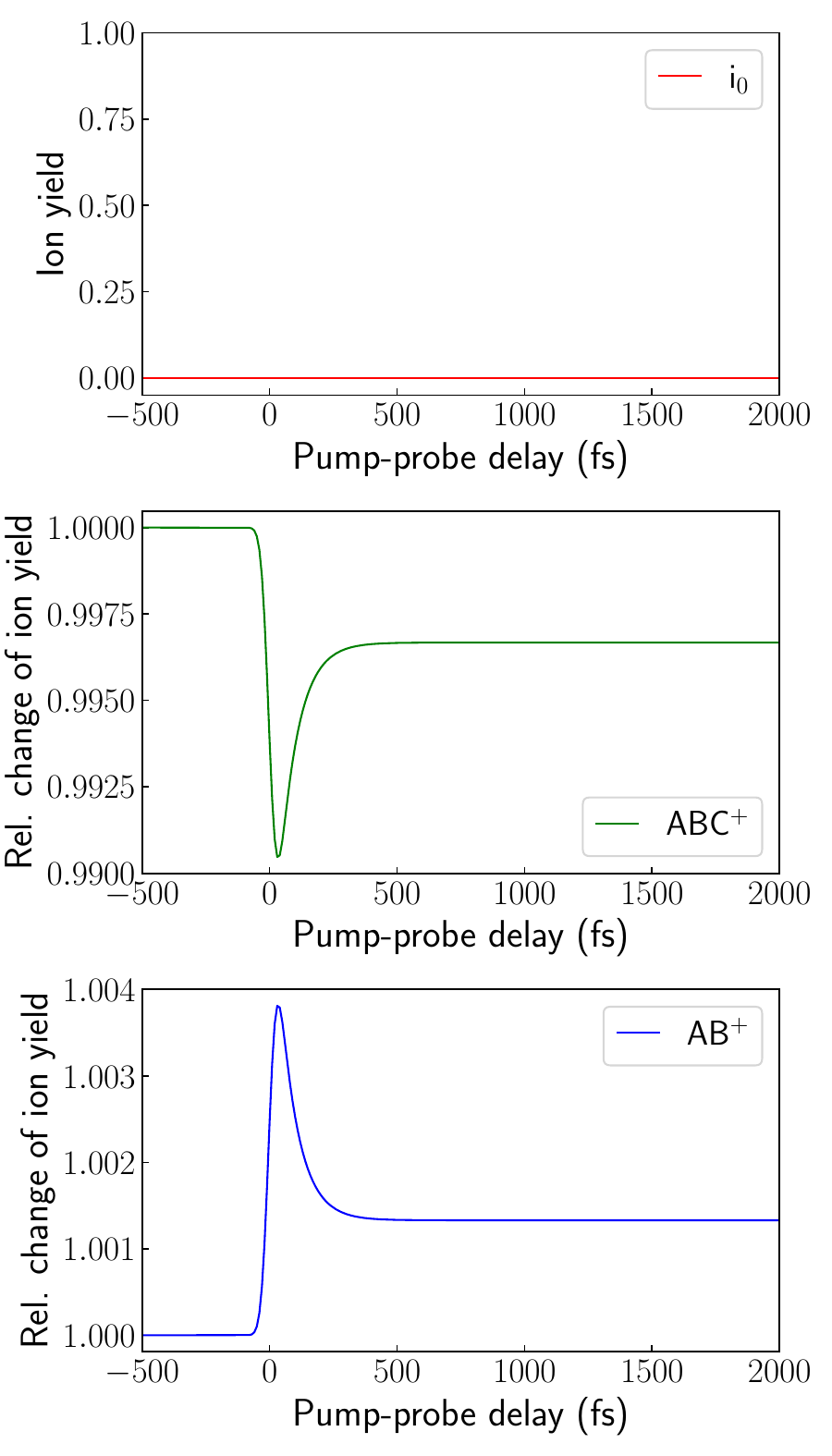}%
   \caption{Simulated disruptive probing measurement in case of Example~2.}
   \label{fig:Model-2}
\end{figure}

As expected, \ip ions show zero population in any pump--probe measurement because the initial
population is transferred to \ABCp and \ABp ions. The other two ion states exhibit relative signal
decreases and increases. This flow of the ion population from one ion fragment type to another is
connected to the slight changes in the rate constants, and we would observe similar pump--probe
signal dependencies except in cases where $\delta_{\ip \rightarrow \ABCp} = \delta_{\ip \rightarrow
\ABp}$.

Both ion signal dependencies are equal to the \ip relative ion yield dependency shown in
\autoref{fig:Dynamics-1}, multiplied by a constant dependent on the perturbed and unperturbed rate
constants. This can be verified by fitting the signal dependencies. \autoref{fig:Fitting-1} shows
the result of fitting the \ABCp signal. The fitted rate constant is $k^{fit} = 1/(71.4~\text{fs})$,
which exactly matches $k_{\ip} = k_{\ip \rightarrow \ABCp} + k_{\ip \rightarrow \ABp}$. The same
rate constant would be obtained from fitting the change in \ABp ion yield. Thus, from this simple
example, we see that qualitative analysis of the disruptive-probing pump--probe signal dependencies
provides the rate constant (or lifetime) for the decay of \ip. To obtain the individual rate
constants $k_{\ip \rightarrow \ABCp}$ and $k_{\ip \rightarrow \ABp}$, additional information is
needed, such as the ratio of \ABCp and \ABp ion yields in undisturbed measurements.

Example 1 was the simplest possible model scenario illustrating the disruptive effect in a
pump--probe measurement using the disruptive probing technique. The next six examples will introduce
additional features that can provide a deeper understanding of the origin of the observed signal
dependencies.

\subsection{Kinetic model: Example 2}
\label{sec:disprob-example2}

\begin{figure}
   \includegraphics[width=\linewidth]{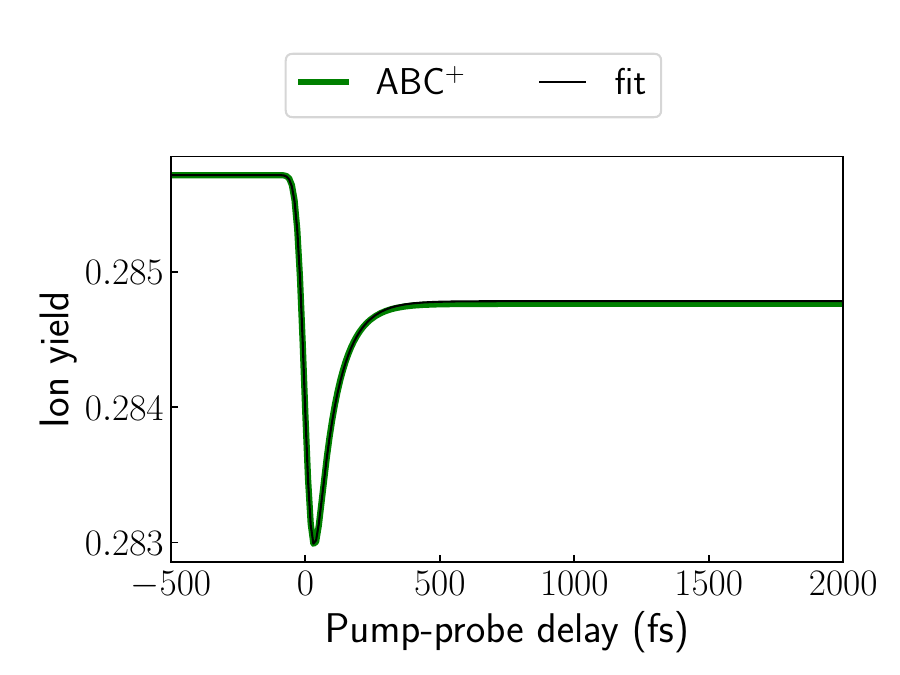}%
   \caption{Fitted simulated signal dependence of ABC$^+$ ion from Example~2.}
   \label{fig:Fitting-2}
\end{figure}

Example 2 illustrates a situation similar to that described in Example 1, using the same 3-level
system. In this case, the probing pulse induces an additional coupling $k\delta_{\ABCp \rightarrow
\ABp}h(t)$ between \ABCp and \ABp ions, resulting in additional fragmentation of \ABCp. The set of
differential equations \ref{eq:m1-ABC} and \ref{eq:m1-AB} describing this situation is modified by
including an additional term $k\delta_{\ABCp \rightarrow \ABp} = 1/(300~\text{fs})$, as follows:
\begin{align}
   \begin{split} \odv{[\ABCp]}{t} =& + k_{\text{i}_0\rightarrow\ABCp}(1 + \delta_{\ip\rightarrow
\ABCp} h(t)) [\ip] \\ & - k\delta_{\ABCp\rightarrow\ABp} h\left(t\right) [\ABCp], \label{eq:m2-ABC}
   \end{split} \\\
   \begin{split} \odv{[\ABp]}{t} =& + k_{\text{i}_0\rightarrow\ABp} (1 + \delta_{\ip\rightarrow
\ABCp} h(t)) [\ip] \\ & + k\delta_{\ABCp\rightarrow\ABp} h\left(t\right) [\ABCp]. \label{eq:m2-AB}
   \end{split}
\end{align}

The new coupling rate constant is denoted as $k\delta_{\text{X}^+ \rightarrow \text{Y}^+}$ to
indicate that it is induced by the probing laser field represented by $h\left(t\right)$. The
simulation of the disruptive probing measurement under these conditions is shown in
\autoref{fig:Model-2}.

A lasting signal step is observed, reflecting that the probing laser-induced couplings transfer part
of the \ABCp population to the \ABp ion yield. Notice that this transfer is independent of the
pump--probe delay due to the additional energy, which destabilizes the \ABCp ions. The time
independence of this feature enables us to distinguish this example from the disruption of an
ongoing dynamical process. The question remains if we can retrieve the decay rate of \ip.

By fitting the pump--probe dependency of \ABCp with a model function consisting of an
exponentially-modified-Gaussian function (\autoref{eq:exp-gauss}) combined with a step function $A
\cdot [1 + \text{erf}(\frac{x-\mu}{\sigma})]/2$, we obtain $k^{fit} = 1/(71.1~\text{fs})$ as an
estimate of the \ip decay rate constant. The fit is shown in \autoref{fig:Fitting-2}. We would
obtain the same result for the dependence of the \ABp ion signal, but with the opposite sign of the
amplitude. The fitted rate constant is slightly (less than 1\%) different due to the additional
coupling; nevertheless, we still achieve good agreement with the simulated decay rate of \ip.

Example~2 illustrates that even the additional disruption of the finally detected ions should not
prevent the retrieval of the information about the original fragmentation dynamics.

\subsection{Kinetic model: Example 3}
\label{sec:disprob-example3}

\begin{figure}
   \includegraphics[width=\linewidth]{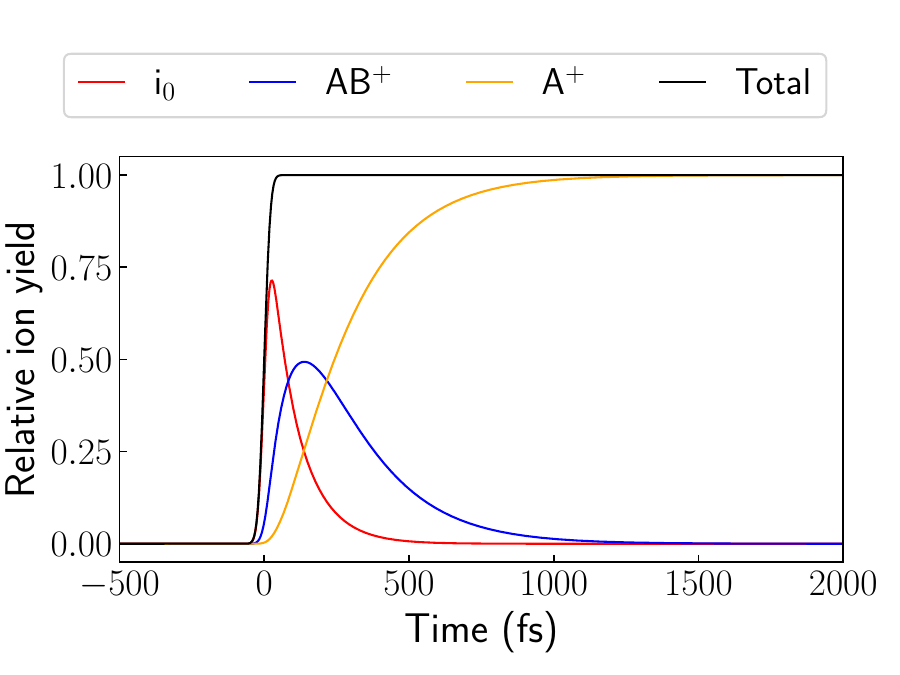}%
   \caption{Dynamics of the ABC compound after ionization according to the Example~3.}
   \label{fig:Dynamics-3}
\end{figure}

\begin{figure}
   \includegraphics[width=\linewidth]{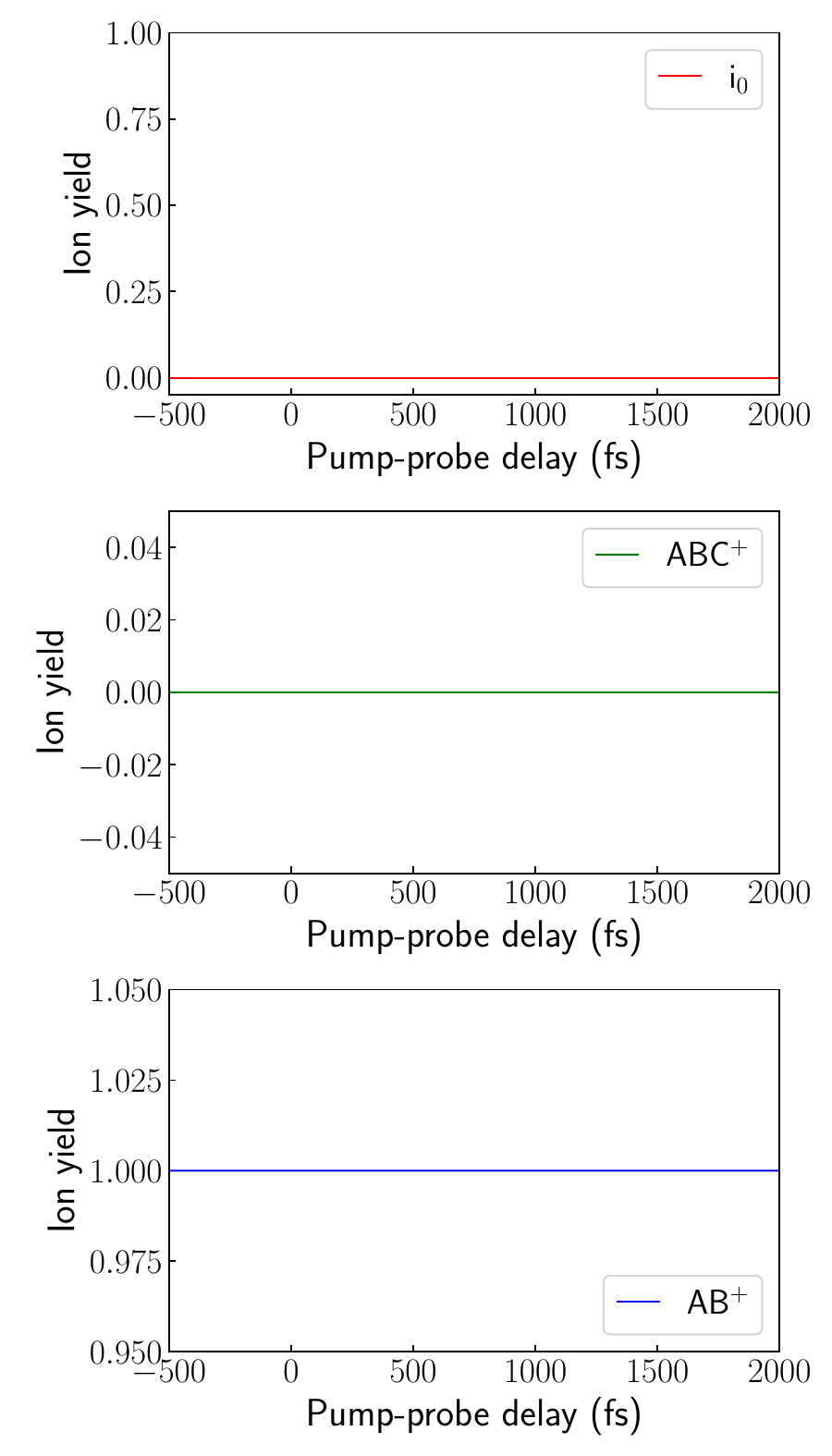}%
   \caption{Simulated disruptive probing measurement in case of Example~3.}
   \label{fig:Model-3}
\end{figure}

The next example of a disruptive probing measurement will be demonstrated using a 3-ionic-state
system with sequential decay. The corresponding set of differential equations is written as follows:
\begin{align}
   \begin{split} \odv{[\ip]}{t} =& g(t) \\ & - k_{\text{i}_0\rightarrow\ABp} (1 +
\delta_{\ip\rightarrow \ABp}h\left(t\right))[\ip], \label{eq:m3-ip}
   \end{split} \\
   \begin{split} \odv{[\ABp]}{t} =& + k_{\text{i}_0\rightarrow\ABp}(1 + \delta_{\ip\rightarrow
\ABp}h\left(t\right))[\ip] \\ & - k_{\ABp\rightarrow\Ap}(1 + \delta_{\ABp\rightarrow
\Ap}h\left(t\right))[\ABp] , \label{eq:m3-AB}
   \end{split} \\ \odv{[\Ap]}{t} =& + k_{\ABp\rightarrow\Ap}(1 + \delta_{\ABp\rightarrow
\Ap}h\left(t\right))[\ABp], \label{eq:m3-A}
\end{align} where we have included the terms induced by the probing pulse, specifically $\delta_{\ip
\rightarrow \ABp} = 2.5$ and $\delta_{\ABp \rightarrow \Ap} = 4$, as well as the rate constant
$k_{\ABp \rightarrow \Ap} = 1 / (200~\text{fs})$. The resulting relative ion signal dependencies of
interest are shown in~\autoref{fig:Dynamics-3}. We can observe that, in the end, the entire
population is transferred to the \Ap fragments. Similarly, if the decay-rate-affecting-probing field
were present, all the population would still end up in the \Ap fragments. Therefore, using
disruptive probing with the mass spectrometric technique is unable to provide any insight into the
ongoing dynamics. This is illustrated in \autoref{fig:Model-3}.

\subsection{Kinetic model: Example 4}
\label{sec:disprob-example4}

\begin{figure}
   \includegraphics[width=\linewidth]{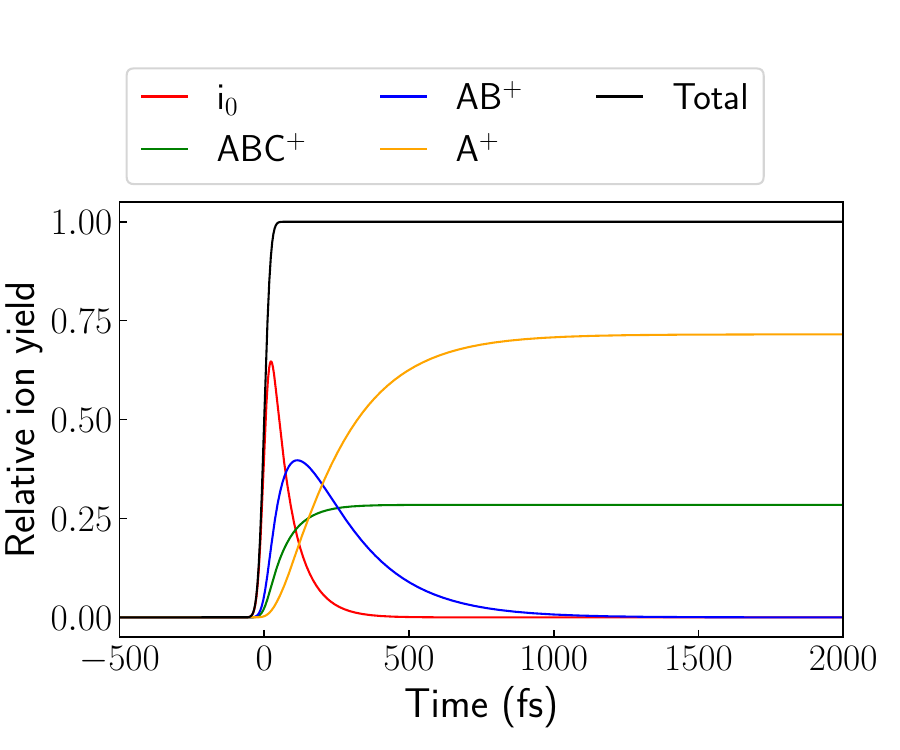}%
   \caption{Dynamics of the ABC compound after ionization according to the Example 4.}
   \label{fig:Dynamics-4}
\end{figure}

\begin{figure}
   \includegraphics[width=\linewidth]{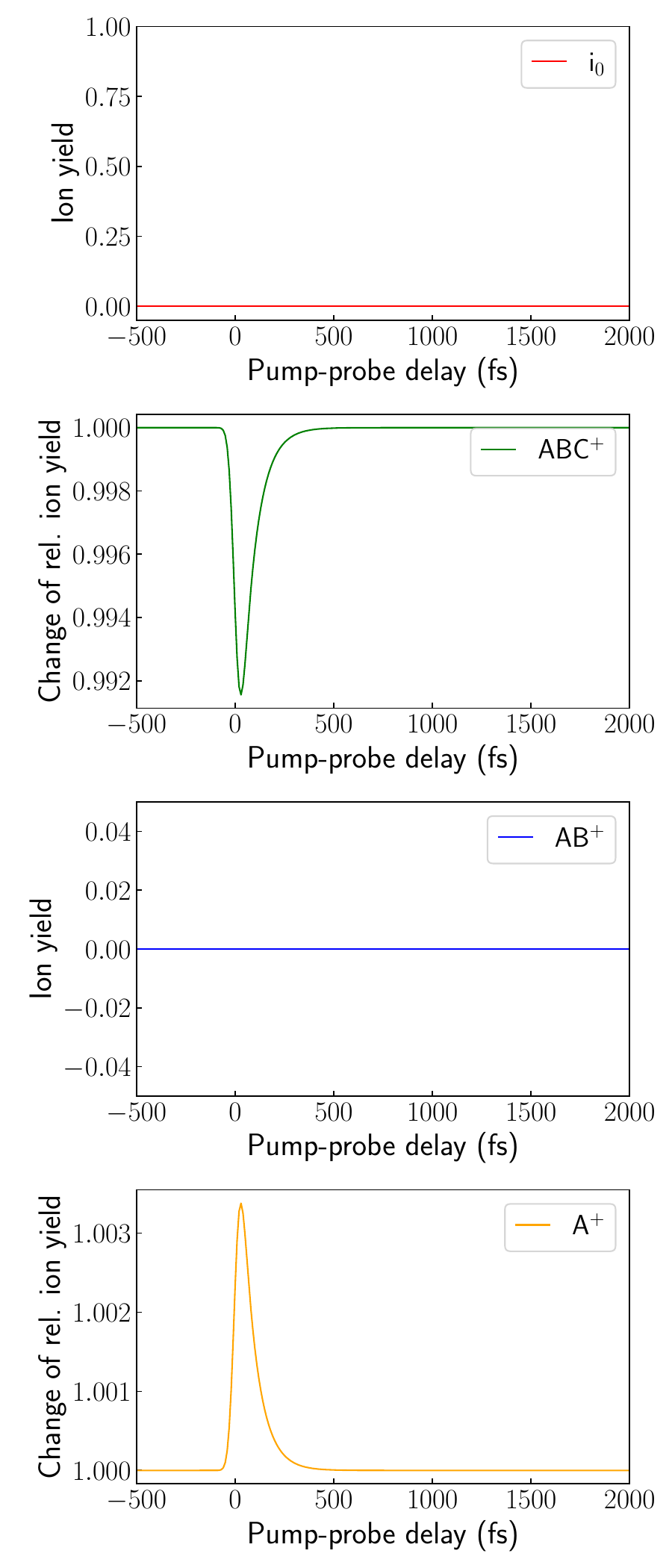}%
   \caption{Simulated disruptive probing measurement in case of Example~4.}
   \label{fig:Model-4}
\end{figure}

Similar to Example~3, Example~4 examines the disruptive probing measurement on a system undergoing
sequential fragmentation: $\ip \rightarrow \ABp \rightarrow \Ap$. However, this time there is also
an \ABCp level present. We apply all the constants used in the previous examples, except for the
coupling between the \ABCp and \ABp levels. The differential equations describing this system,
including the effects of the probing pulse, are as follows:
\begin{align}
   \begin{split} \odv{[\ip]}{t} =& g(t) \\ & - k_{\text{i}_0\rightarrow\ABp} (1 +
\delta_{\ip\rightarrow \ABp}h\left(t\right))[\ip] \\ & - k_{\text{i}_0\rightarrow\ABCp} (1 +
\delta_{\ip\rightarrow \ABCp}h\left(t\right))[\ip] \label{eq:m4-ip}
      \end{split} \\ \odv{[\ABCp]}{t} =& + k_{\text{i}_0\rightarrow\ABCp}(1 + \delta_{\ip\rightarrow
\ABCp}h\left(t\right))[\ip], \label{eq:m4-ABC} \\
   \begin{split} \odv{[\ABp]}{t} =& + k_{\text{i}_0\rightarrow\ABp}(1 + \delta_{\ip\rightarrow
\ABp}h\left(t\right))[\ip] \\ & - k_{\ABp\rightarrow\Ap}(1 + \delta_{\ABp\rightarrow
\Ap}h\left(t\right))[\ABp] , \label{eq:m4-AB}
   \end{split} \\ \odv{[\Ap]}{t} =& + k_{\ABp\rightarrow\Ap}(1 + \delta_{\ABp\rightarrow
\Ap}h\left(t\right))[\ABp]. \label{eq:m4-A}
\end{align} The undisturbed evolution of the states' populations is shown in
\autoref{fig:Dynamics-4}. This time, the resulting population ends either as \Ap ions from the
sequential fragmentation or \ABCp ions due to \ip stabilization. Therefore, the disruptive probing
measurement will reveal changes in the resulting ion yields, as illustrated in the simulated
measurement shown in \autoref{fig:Model-4}.

We can see that the outcome of the measurement is the same as in Example~1. For the given values of
rate constants, the measurement suggests that the additional level \ABp is not detected. The fitted
rate constant, obtained using an exponentially-modified-Gaussian function, is $k^{fit} =
1/(71.44~\text{fs})$, which is similar to the decay rate of \ip.

Example~4 illustrates that in such a case, we are blind to the metastable level, even when an
alternative route is present. The next two examples will present opposite situations, where the
sequential fragmentation dynamics can be retrieved using the disruptive probing method.

\subsection{Kinetic model: Example 5}
\label{sec:disprob-example5}

\begin{figure}
   \includegraphics[width=\linewidth]{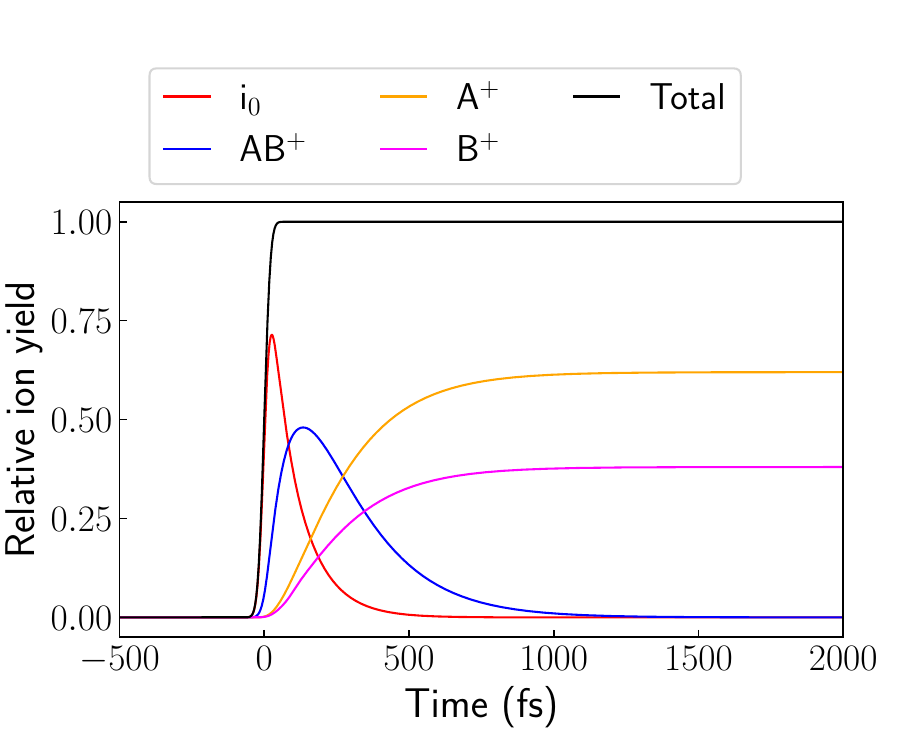}%
   \caption{Dynamics of the ABC compound after ionization according to the Example 5.}
   \label{fig:Dynamics-5}
\end{figure}

\begin{figure}
   \includegraphics[width=\linewidth]{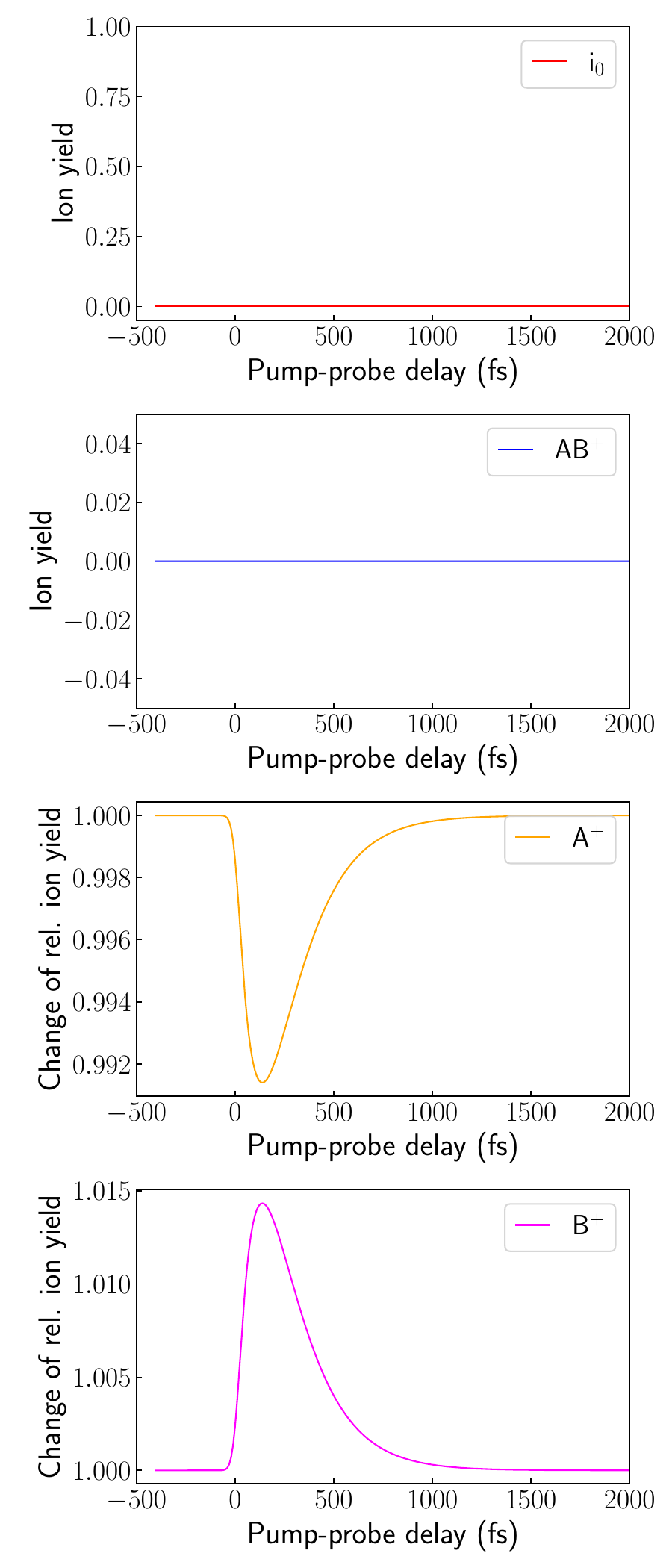}%
   \caption{Simulated disruptive probing measurement in case of Example~5.}
   \label{fig:Model-5}
\end{figure}

\begin{figure}
   \includegraphics[width=\linewidth]{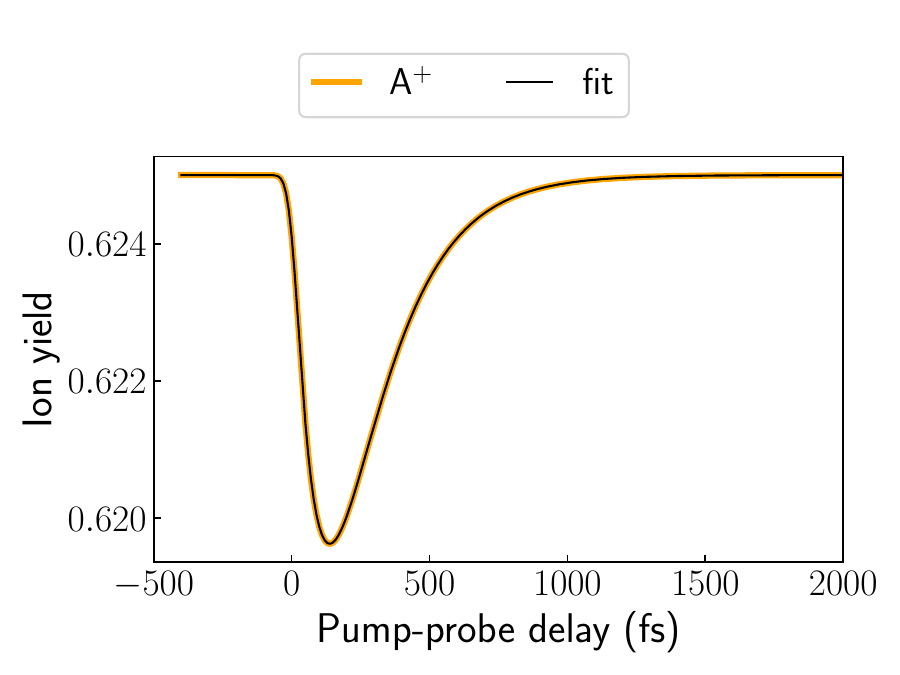}%
   \caption{Fitted simulated measurement of A$^+$ ion from Example~5.}
   \label{fig:Fitting-5}
\end{figure}

The situation changes from the previous example when the two levels result from sequential
fragmentation, such as $\ip \rightarrow \ABp \rightarrow \Ap$ or $\ip \rightarrow \ABp \rightarrow
\Bp$. The new set of differential equations is as follows:
\begin{align}
   \begin{split} \odv{[\ip]}{t} =& g(t) \\ & - k_{\text{i}_0\rightarrow\ABp} (1 +
\delta_{\ip\rightarrow \ABp}h\left(t\right))[\ip] \label{eq:m5-ip}
      \end{split} \\
      \begin{split} \odv{[\ABp]}{t} =& + k_{\text{i}_0\rightarrow\ABp}(1 + \delta_{\ip\rightarrow
\ABp}h\left(t\right))[\ip] \\ & - k_{\ABp\rightarrow\Ap}(1 + \delta_{\ABp\rightarrow
\Ap}h\left(t\right))[\ABp] \\ & - k_{\ABp\rightarrow\Bp}(1 + \delta_{\ABp\rightarrow
\Bp}h\left(t\right))[\ABp] , \label{eq:m5-AB}
      \end{split} \\ \odv{[\Ap]}{t} =& + k_{\ABp\rightarrow\Ap}(1 + \delta_{\ABp\rightarrow
\Ap}h\left(t\right))[\ABp]. \label{eq:m5-A} \\ \odv{[\Bp]}{t} =& + k_{\ABp\rightarrow\Bp}(1 +
\delta_{\ABp\rightarrow \Bp}h\left(t\right))[\ABp], \label{eq:m5-B}
\end{align} where we set the rate constants as $k_{\ip \rightarrow \ABp} = 1/(100~\text{fs})$,
$k_{\ABp \rightarrow \Ap} = 1/(300~\text{fs})$, and $k_{\ABp \rightarrow \Bp} = 1/(500~\text{fs})$.
Additionally, the induced disruption by the probing pulse were set as $\delta_{\ip \rightarrow \ABp}
= 2.5$, $\delta_{\ABp \rightarrow \Ap} = 1$, and $\delta_{\ABp \rightarrow \Bp} = 10$.

By numerically solving the differential equations, one can obtain the ion signal dependencies for
the undisturbed case, as shown in \autoref{fig:Dynamics-5}, and the simulated result of the
pump--probe disruptive probing measurement, illustrated in \autoref{fig:Model-5}. The rate fitting
using two exponentially modified Gaussian functions, as shown in \autoref{fig:Fitting-5}, reveals
the rate constants for the population and decay of the \ABp ion level. We obtain $k^{fit}_1 =
1/(100~\text{fs})$ and $k^{fit}_2 = 1/(187~\text{fs})$, which correspond to $k_{\ip \rightarrow
\ABp}$ and $k_{\ABp \rightarrow \Ap} + k_{\ABp \rightarrow \Bp}$, respectively.

The sequential pathway splitting into final fragments \Ap and \Bp enables us to detect the rate
constant for population and following fragmentation of the metastable \ABp ion. This is very similar
to Example~1, where the population rate was given by a laser pulse time-dependence.

\subsection{Kinetic model: Example 6}
\label{sec:disprob-example6}

\begin{figure}
   \includegraphics[width=\linewidth]{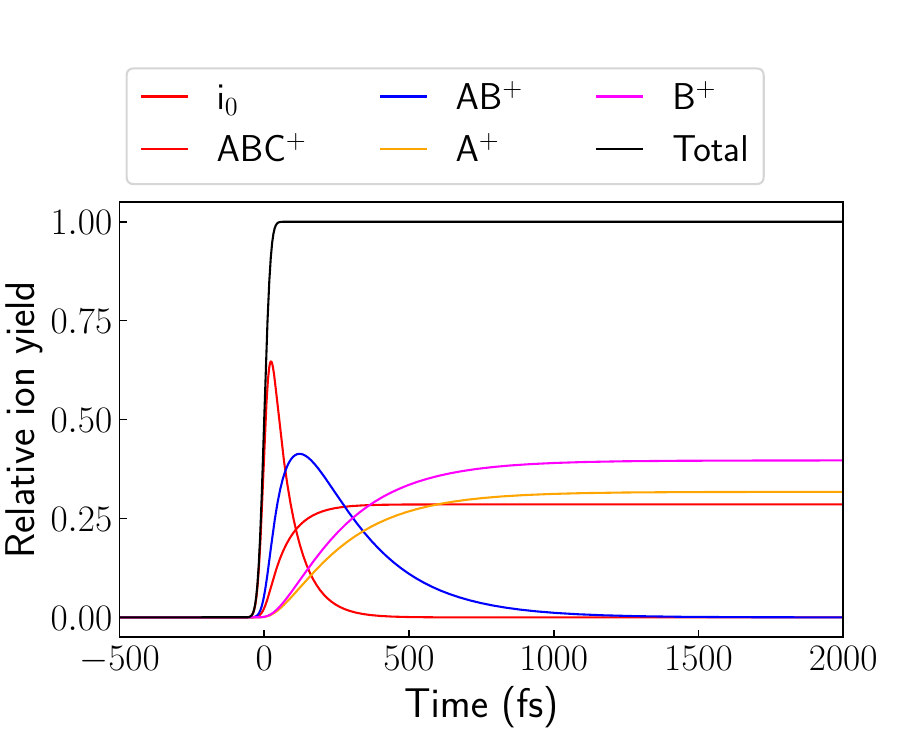}%
   \caption{Dynamics of the ABC compound after ionization according to the Example~6.}
   \label{fig:Dynamics-6}
\end{figure}

\begin{figure}
   \includegraphics[width=\linewidth]{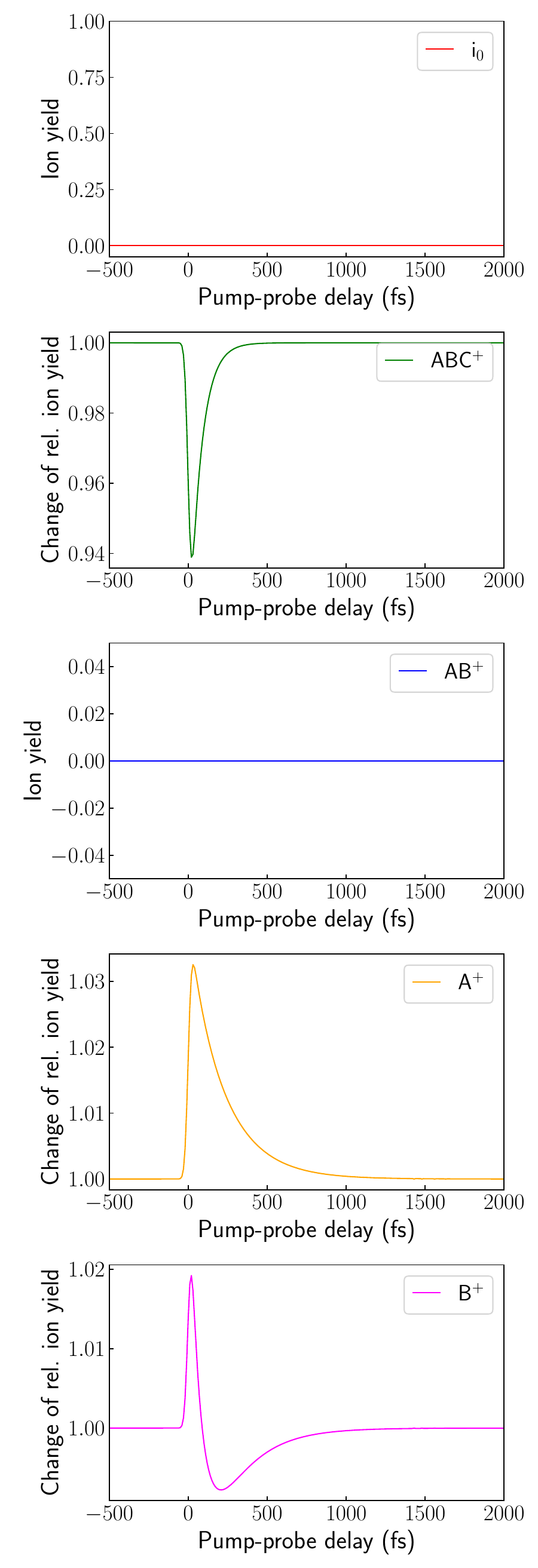}%
   \caption{Simulated disruptive probing measurement in case of Example~6.}
   \label{fig:Model-6}
\end{figure}

\begin{figure}
   \includegraphics[width=\linewidth]{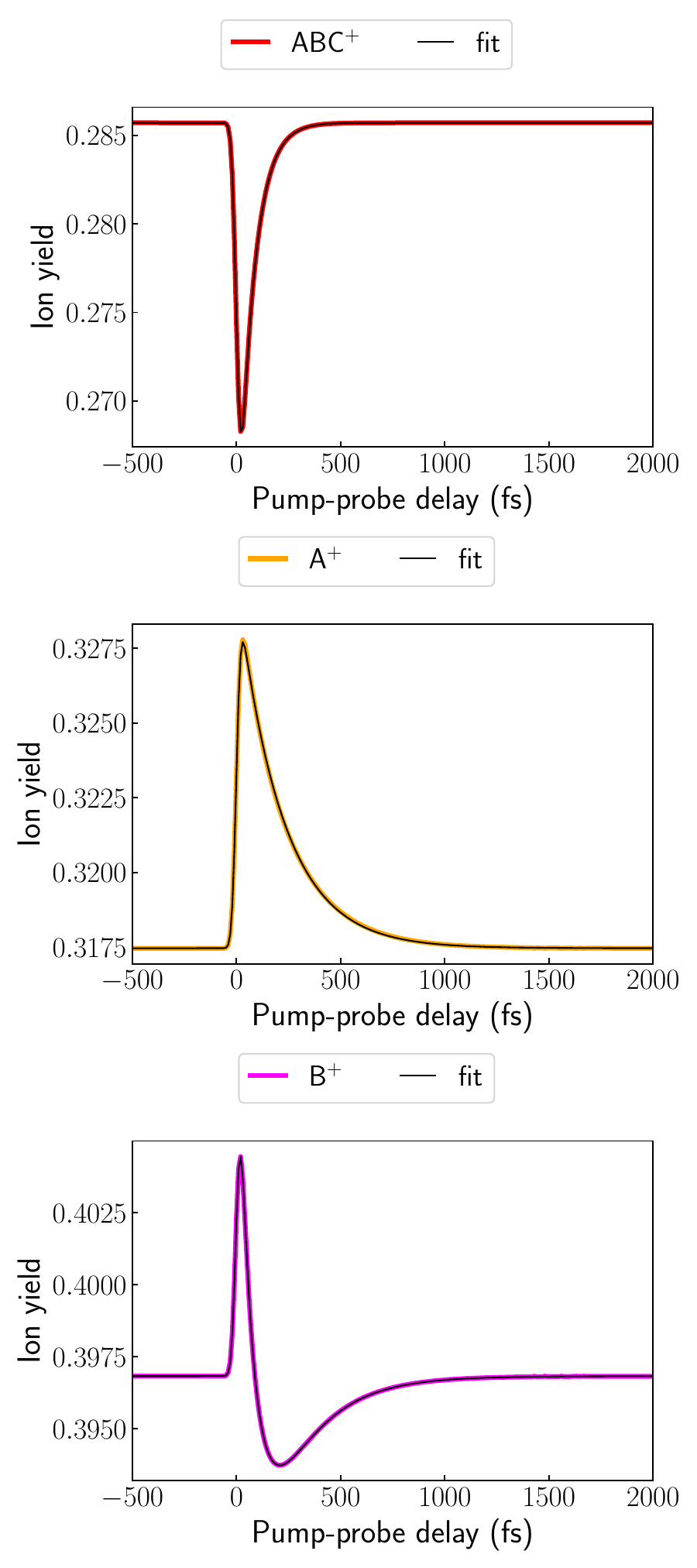}%
   \caption{Fitted simulated signal dependences of ABC$^+$, A$^+$, and B$^+$ ion from Example~6.}
   \label{fig:Fitting-6}
\end{figure}

The most complex example incorporates all the previous cases, with the addition of the stabilized
\ABCp as an alternative pathway for the populated \ip state. The dynamics of the five ionic levels,
denoted as \ip, \ABCp, \ABp, \Ap, and \Bp, is described by the following variables:
\begin{align}
   \begin{split} \odv{[\ip]}{t} =& g(t) \\ & - k_{\text{i}_0\rightarrow\ABp} (1 +
\delta_{\ip\rightarrow \ABp}h\left(t\right))[\ip] \\ & - k_{\text{i}_0\rightarrow\ABCp}
[\ip] \label{eq:m6-ip}
      \end{split} \\ \odv{[\ABCp]}{t} =& + k_{\text{i}_0\rightarrow\ABCp}[\ip], \label{eq:m6-ABC} \\
      \begin{split} \odv{[\ABp]}{t} =& + k_{\text{i}_0\rightarrow\ABp}(1 + \delta_{\ip\rightarrow
\ABp}h\left(t\right))[\ip] \\ & - k_{\ABp\rightarrow\Ap}(1 + \delta_{\ABp\rightarrow
\Ap}h\left(t\right))[\ABp] \\ & - k_{\ABp\rightarrow\Bp}[\ABp] , \label{eq:m6-AB}
      \end{split} \\ \odv{[\Ap]}{t} =& + k_{\ABp\rightarrow\Ap}(1 + \delta_{\ABp\rightarrow
\Ap}h\left(t\right))[\ABp]. \label{eq:m6-A} \\ \odv{[\Bp]}{t} =& +
k_{\ABp\rightarrow\Bp}[\ABp], \label{eq:m6-B}
\end{align} where original rate constants are $1/k_{\text{i}_0\rightarrow\ABCp} = 250$~fs,
$1/k_{\text{i}_0\rightarrow\ABCp} = 100$~fs, $1/k_{\ABp\rightarrow\Ap} = 500$~fs,
$1/k_{\ABp\rightarrow\Ap} = 400$~fs, and induced disruptive changes are $\delta_{\ip\rightarrow
\ABp} = \delta_{\ABp\rightarrow \Ap} = 10$. The resulting undisturbed dynamics are shown in
\autoref{fig:Dynamics-6}. Clearly, the population of each level develops differently.

The simulated measurement result is shown in \autoref{fig:Model-6}. All four detected ions, \ie,
\ABCp, \ABp, \Ap, and \Bp, exhibit different signal dependence in the simulated disruptive probing
measurement. Specifically, the \ABCp ion signal reveals a narrow negative
exponentially-modified-Gaussian dependence, similar to the dynamics of the \ip signal, while the \Ap
and \Bp ion signal dependences are linked to the population and decay of the \ABp ions.

This enables us to obtain three different rate constants just from the data fitting, as shown in
\autoref{fig:Fitting-6}. The \ABCp signal was fitted with a single exponentially-modified-Gaussian
function, while the \Ap and \Bp signals were fitted with two exponentially-modified-Gaussian
functions. The resulting fits are shown in~\autoref{fig:Fitting-6}. The obtained rates of
$1/k^\text{fit}_{\ABCp} = 71.4$~fs, $1/k^\text{fit}_{1\Ap} = 71.4$~fs, and $1/k^\text{fit}_{2\Ap} =
222.2$~fs are in perfect agreement with the used rate constants for the dynamics simulation.

\begin{figure}
   \includegraphics[width=\linewidth]{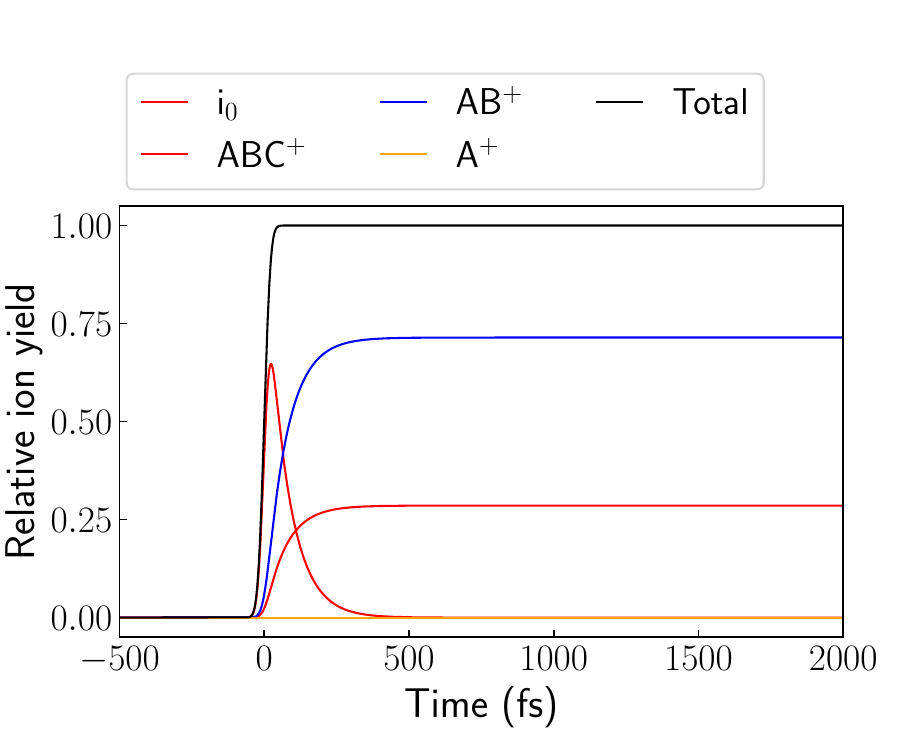}%
   \caption{Dynamics of the ABC compound after ionization according to the Example~7.}
   \label{fig:Dynamics-7}
\end{figure}

\begin{figure}
   \includegraphics[width=\linewidth]{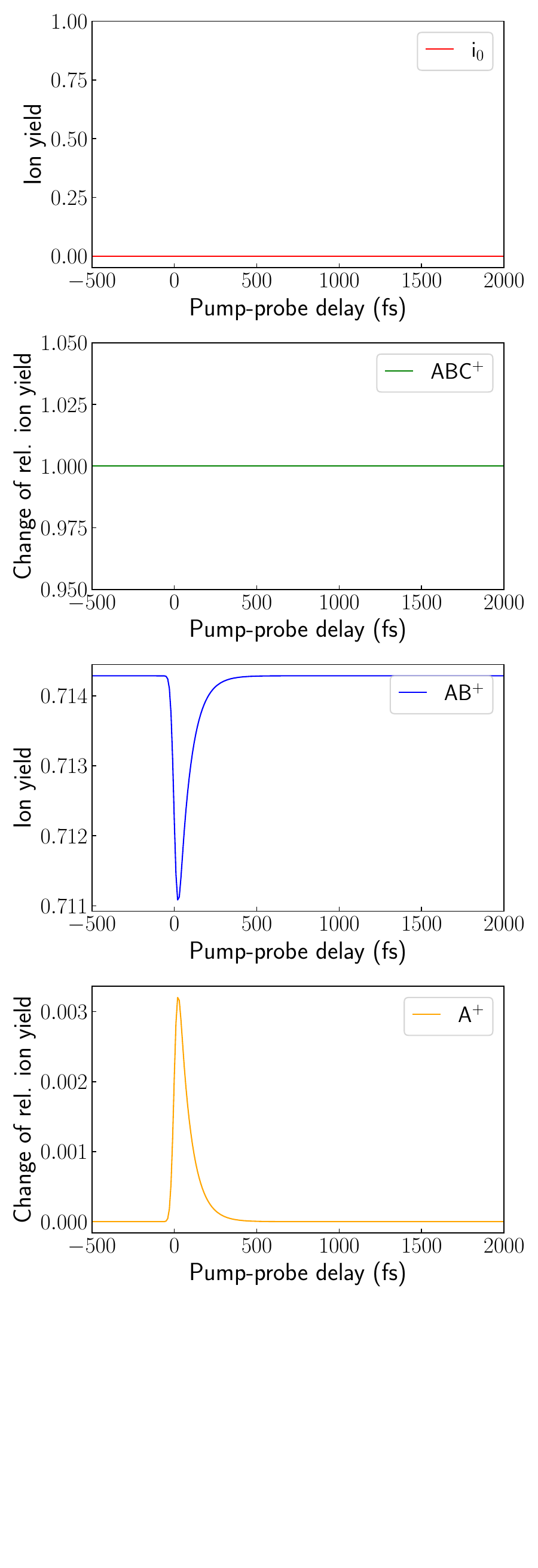}%
   \caption{Simulated disruptive probing measurement in case of Example~7.}
   \label{fig:Model-7}
\end{figure}

\begin{figure}
   \includegraphics[width=\linewidth]{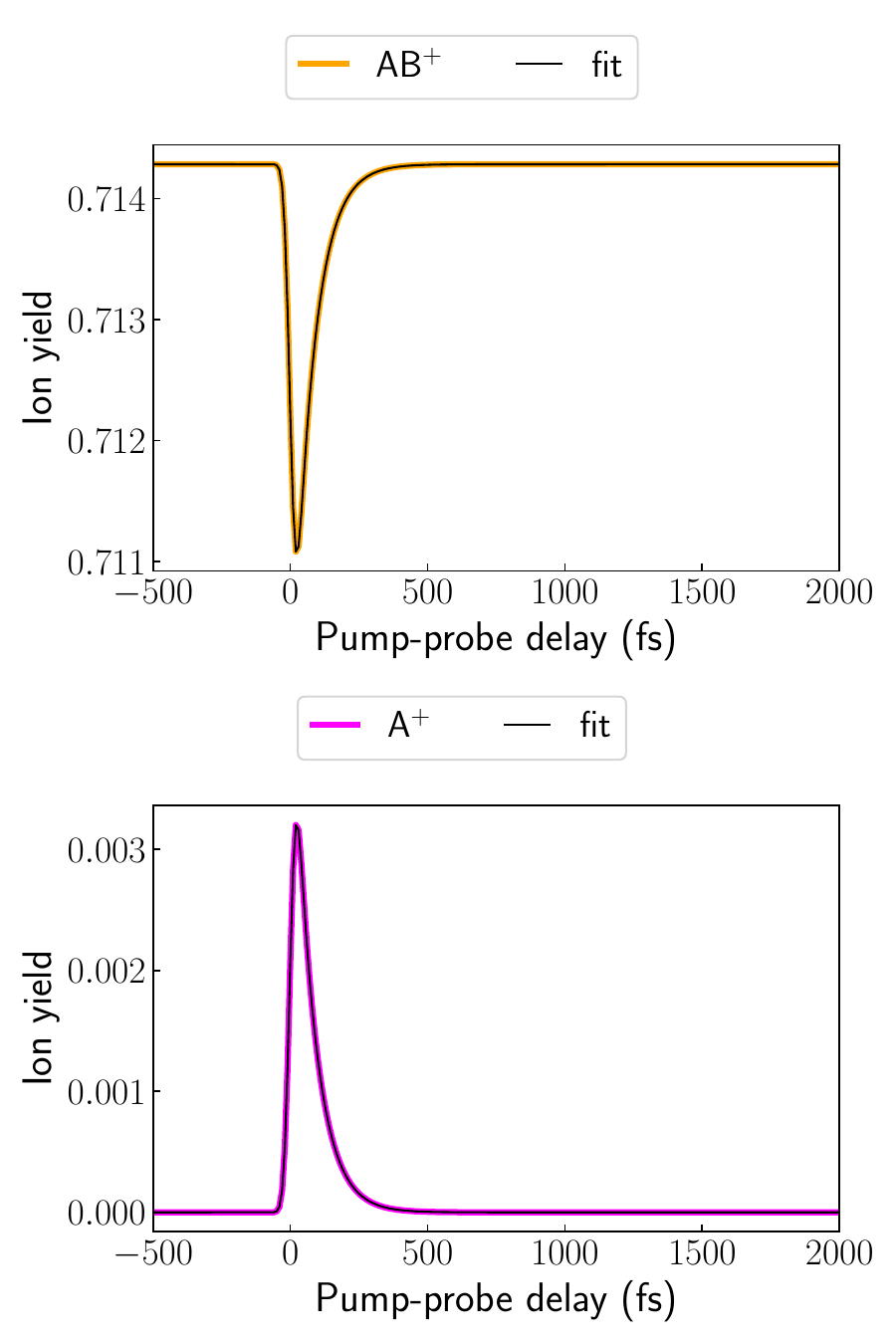}%
   \caption{Fitted simulated signal dependences of AB$^+$, and A$^+$ ion from Example~7.}
   \label{fig:Fitting-7}
\end{figure}

\subsection{Kinetic model: Example 7}

The final example shows a different scenario of the disruptive effect. Instead of reshuffling the
ions' final population among the existing pathways, like in Examples~1 and 6, or reshuffling the ion
signal among the final ion species, like in Example~2, the disrupting pulse opens of a new
fragmentation pathway leading to detection of a new ion species. This can be described in
differential equations:
\begin{align}
    \begin{split} \odv{[\ip]}{t} =& g(t) \\ & - k_{\text{i}_0\rightarrow\ABCp} [\ip] -
k_{\text{i}_0\rightarrow\ABp} ,[\ip] \label{eq:m7-ip}
    \end{split} \\
    \begin{split} \odv{[\ABCp]}{t} =& + k_{\text{i}_0\rightarrow\ABCp}[\ip], \label{eq:m7-ABC}
    \end{split} \\ \odv{[\ABp]}{t} =& + k_{\ip\rightarrow\ABp}(1 - r_{\ABp\leftrightarrow\Ap}
h\left(t\right))[\ip], \label{eq:m7-AB} \\ \odv{[\Ap]}{t} =& +
k_{\ip\rightarrow\ABp}(r_{\ABp\leftrightarrow\Ap} h\left(t\right))[\ip]. \label{eq:m7-A}
\end{align}

These equations describe the dynamics shown in~\autoref{fig:Dynamics-7}. Notice that in the absence
of the perturbing pulse, the \Ap signal is zero. This is changed by the presence of the disrupting
pulse as is plotted in~\autoref{fig:Model-7}, illustrating simulation of disruptive probing
measurement. The fitting of the \ABp and \Ap pump--probe signal presented in~\autoref{fig:Fitting-7}
reveals the fragmentation rates corresponding to the \ip dynamics. Importantly, the \ABCp channel is
not affected.

Although this is only a model situation for a disruptive probing measurement, it also illustrates
that the signals of different ion species, which the pump--probe signal shows a correlation, do not
necessarily correspond to the topology of the original dynamics. In contrast, the described model
scenarios support that the dynamical observations revealed by the disruptive probing technique are
inherently connected to the original dynamics of the system.

\subsection{Summary}
\label{sec:disprob-summary}
\begin{figure*} 
   \includegraphics[width=1\linewidth]{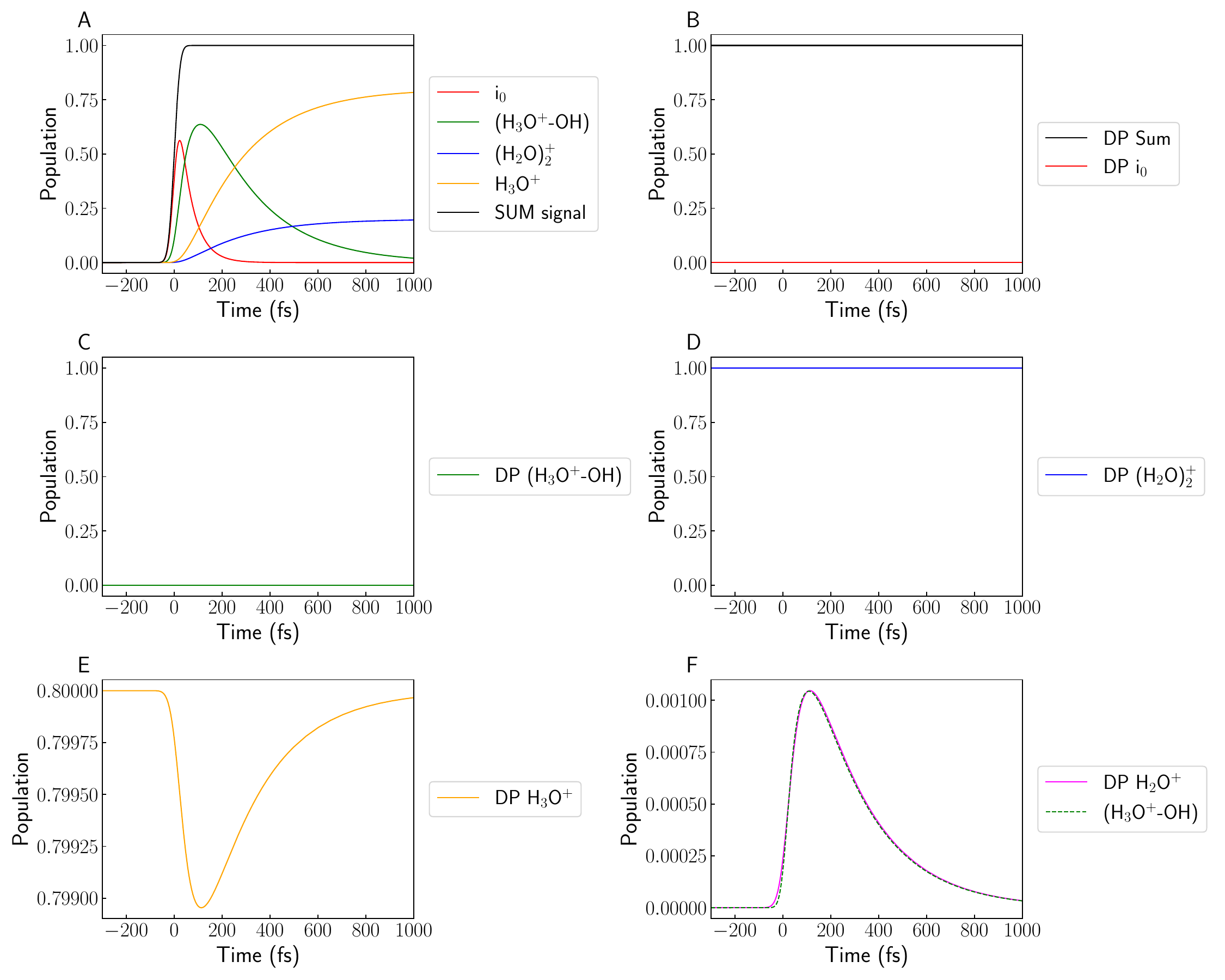}%
   \caption{Numerical simulation of fragmentation of singly ionized water dimer dynamics from the
      ground and the first excited states using kinetics model described
      by~\autoref{eq:dp-w2-1}-\autoref{eq:dp-w2-5}. (A) Unperturbed dynamics. (B-F) Simulation of
      the pump--probe signal dependencies in disruptive probing experiment for different ion
      species. (F) A comparison between the searched dynamics of $[\HHHOp{}\cdots\text{OH}]$ (dashed
      green) with the simulated pump--probe signal dependence of \HHOp.}
   \label{fig:DP-simulation-W2}
\end{figure*}
The simple examples discussed illustrate how pump--probe disruptive probing measurements can yield
markedly different results, depending on the topology of the system under study and the nature of
the induced coupling changes.

In simple terms, the ansatz based on differential equations for describing disruptive probing
experiments models the probe-induced, time-dependent changes in the ongoing dynamics. These changes
can manifest as enhancements or diminutions of rate constants, the induction of new reaction
pathways, or fragmentation into new ion products. Various mechanisms may be responsible for these
effects, such as Stark coupling, rovibrational excitation, or Raman transitions. Accurately
describing these processes in detail is extremely challenging and currently beyond our capabilities.

Nevertheless, the observed dynamics in pump--probe signals can provide a broad overview of the
ongoing, unperturbed fragmentation processes. The models presented here can be particularly useful
for an initial interpretation of pump--probe data and for preliminary discussion before engaging in
more advanced molecular dynamics simulations.

\subsection{Kinetic simulation of the disruptive probing experiment}
\label{sec:DP-W2-simulation}
The fitting of the pump--probe data shown in~Figure~3 in the main text reveals three significant
dynamics features occuring on different timescales. Firstly, fast process with effective lifetime in
the order of tenths of femtoseconds revealed for \HHHOp, \HHOp. Secondly, the dynamics occuring with
effective lifetime between 200-400~fs for \HHHOp, \HHOp, OH$^+$, and O$^+$. And lastly, slow
dynamics of \HHOdp and H$^+$ in picosecond timescale.

Under our experimental condition, we can expect that the most of the singly ionized \HHOd are
occupying ground and the first excited state. In paper by Svoboda \etal~\cite{Svoboda:PCCP15:11531},
their simulation revealed that singly ionized water dimer in the first two electronic states should
firstly proceed through proton transfer in tenths of femtoseconds leading to the creation of
ion-radical pair $[\HHHOp{}\cdots\text{OH}]$. Subsequently, this is either stabilized or leads to \HHHOp +
OH fragmentation on the timescale of hundreds of femtoseconds.

The observed timescales of the first two features in the experimental data are in good agreement
with the reported theory~\cite{Svoboda:PCCP15:11531}. On the contrary, we would expect similar
features observed not only in the \HHHOp, \HHOp, OH$^+$, and O$^+$ pump--probe signals, but also in
\HHOdp, which is not the case. Here we describe a model of disruptive probing experiment, which
should explain the discrepancy.

The dynamics according to the Svoboda \etal is shown in~\autoref[A]{fig:DP-simulation-W2}. The
diferential equtions describing the scenario are written as
\begin{align}
   \begin{split} \odv{[\ip]}{t} =& g(t) - k_{\ip \rightarrow \HHHOp - \text{OH}} \cdot
[\ip], \label{eq:dp-w2-1} \\
   \end{split} \\
   \begin{split} \odv{[\HHHOp - \text{OH}]}{t} =& + k_{\ip \rightarrow \HHHOp - \text{OH}} \cdot
[i_0] \\ & - k_{\HHHOp - \text{OH}\rightarrow \HHHOp } \cdot [\HHHOp - \text{OH}] \\ & - k_{\HHHOp -
\text{OH}\rightarrow \HHOdp } \cdot [\HHHOp - \text{OH}], \label{eq:dp-w2-2}
   \end{split} \\ \odv{[\HHOdp]}{t} =& + k_{\HHHOp - \text{OH}\rightarrow \HHOdp } \cdot [\HHHOp -
\text{OH}], \label{eq:dp-w2-3} \\ \odv{[\HHHOp]}{t} =& + k_{\HHHOp - \text{OH}\rightarrow \HHHOp }
\cdot [\HHHOp - \text{OH}]. \label{eq:dp-w2-4}
\end{align}

As it is discussed in the disruptive probing example cases described
in~\autoref{sec:disprob-motivation}, no probe-pulse-induced enhancement, reduction of rate constant
in the can lead to our experimental observation, \ie, no observation of the changes in the \HHOdp
signal and activity in the \HHOp, OH$^+$, and O$^+$ ion signal. Therefore, we must assume that the
probing pulse is in fact causing a transition of the part of the $[\HHHOp{}\cdots\text{OH}]$ particles,
which are ongoing the dissociating pathway into the \HHHOp signal (for the simplicity of the model
we will omit the OH$^+$ and O$^+$ signal dependencies). This can be modelled via introducing a new
reshuffling factor $r_{\HHHOp \leftrightarrow \HHOp}$ in~\autoref{eq:dp-w2-4} and adding a new
equation for induced \HHOp population as
\begin{align}
   \begin{split} \odv{[\HHHOp]}{t} =& + k_{\HHHOp - \text{OH}\rightarrow \HHHOp} \cdot \left( 1 -
r_{\HHHOp \leftrightarrow \HHOp} \right) \times \\ & \times [\HHHOp -
\text{OH}], \label{eq:dp-w2-4b}
   \end{split} \\
   \begin{split} \odv{[\HHOp]}{t} =& + k_{\HHHOp - \text{OH}\rightarrow \HHHOp } \cdot r_{\HHHOp
\leftrightarrow \HHOp} \times \\ & \times [\HHHOp - \text{OH}]. \label{eq:dp-w2-5}
   \end{split}
\end{align}

In~\autoref[B-F]{fig:DP-simulation-W2}, there are simulated resulting ion signal dependencies for
disruptive probing experiment. We can observe that the dynamics in \HHHOp and \HHOp are mutually
compensating as the probing-pulse does not populate any ions and only transfers population from one
to the other ion species. Interestingly, the applied model links our experimental observations with
the simulations and support our assignment of the dynamics on the scale of tenths and hundreds of
femtoseconds to population and subsequent fragmentation, respectively, of $[\HHHOp{}\cdots\text{OH}]$ as is
illustrated also in \autoref[F]{fig:DP-simulation-W2}.

\onecolumngrid%
\clearpage
\twocolumngrid%
\renewcommand{\bibsection}{}
\bibliography{string,cmi}%

\onecolumngrid%
\listofnotes%